\documentclass[prd,aps,floatfix, superscriptaddress, nofootinbib, twocolumn, English]{revtex4}

\usepackage{amssymb,amsmath}
\usepackage{graphicx}
\usepackage[percent]{overpic}
\usepackage{overpic}
\usepackage{color}
\usepackage[colorlinks=true]{hyperref}
\usepackage{braket}

\def\ba{\begin{eqnarray}}
\def\ea{\end{eqnarray}}

\newcommand{\I}{\text{I}}
\newcommand{\e}{\epsilon}
\def\d{{\rm d}}
\def\M{{\cal M}}

\begin{document}

\title{Dynamic Signatures of Black Hole Binaries with Superradiant Clouds} 

\author{Jun Zhang}
\email{jun.zhang@imperial.ac.uk}
\affiliation{Theoretical Physics, Blackett Laboratory, Imperial College, London, SW7 2AZ, U.K. }
\author{Huan Yang}
\email{hyang@perimeterinstitute.ca}
\affiliation{Perimeter Institute for Theoretical Physics, Waterloo, Ontario N2L 2Y5, Canada }
\affiliation{University of Guelph, Guelph, Ontario N2L 3G1, Canada}

\begin{abstract} 

Superradiant clouds may develop around a rotating black hole, if there is a bosonic field with Compton wavelength comparable to the size of the black hole. In this paper, we investigate the effects of the cloud on the orbits of nearby compact objects. In particular, we consider the dynamical friction and the backreaction due to level mixing. Under these interactions, the probability of a black hole dynamically capturing other compact objects, such as stellar mass black holes and neutron stars, is generally enhanced with the presence of the cloud. For extreme mass ratio inspirals and binary stellar mass binary black holes, the cloud-induced orbital modulation may be detected by observing the gravitational waveform using space borne gravitational wave detectors, such as LISA. Interestingly within certain range of boson Compton wavelength, the enhanced capture rate of stellar mass black holes could accelerate hierarchical mergers, with higher-generation merger product being more massive than the mass threshold predicted by supernova pair instability. These observational signatures provide promising ways of searching light bosons with gravitational waves.

\end{abstract}

\maketitle 

\section{Introduction}

The existence of light bosons have been motivated from various theoretical considerations. For example, QCD axions are proposed as an extension of the standard model which naturally solves the strong CP problem in particle physics \cite{PhysRevLett.38.1440,PhysRevLett.40.223}. It is also shown that light bosons can form a Bose-Einstein condensate in galaxies and could be a good candidate for dark matter \cite{PhysRevD.28.1243, PhysRevLett.64.1084, Hu:2000ke, Peebles:2000yy, Amendola:2005ad, Schive:2014dra, Hui:2016ltb}. 
Moreover, light bosons are predicted in string theory \cite{Marsh:2015xka, arvanitaki2010string}. Scalar degrees of freedom could arise as products of compactifying extra dimensions \cite{Green:1987sp, Svrcek:2006yi}, the mass of which could be in a wide range with a lower bound possibly down to the Hubble scale. Therefore, the detection of light bosons not only provides a smoking gun to new physics, but also has profound indications on string theory. The coupling between these light bosons and normal matter is model dependent, but is weak in general, which makes the detection of such light bosons a difficult task. Nevertheless, thanks to the equivalence principle, these light bosons at least couple to matter gravitationally, which provides a universal way for searching.

Depending on their mass and coupling to the standard model particles, light bosons could lead to different observable effects, including \cite{arvanitaki2010string, Arvanitaki:2010sy, Marsh:2015xka, Baumann:2016wac, Rosa:2017ury, Sagunski:2017nzb, Fujita:2018zaj, Huang:2018pbu,Ikeda:2019fvj}. In particular, the detection of gravitational waves (GWs) made by LIGO and Virgo opened up an new observational window. Together with space borne GW detectors, such as LISA (Laser Interferometer Space Antenna), GW observations provide promising new ways of searching for light bosons. In fact, it has been suggested for a long time that a massive bosonic field near a rotating black hole (BH) may grow exponentially by extracting angular momentum from the BH, a process known as supperradiance \cite{Detweiler:1980uk, zouros1979instabilities, Brito:2015oca, Dolan:2007mj}. The growth of the field can be very efficient if the Compton wavelength of the field is comparable to the size of a rotating BH. More precisely, the growth rate of a scalar field scales as \cite{Detweiler:1980uk}
\ba
\Gamma_{n\ell m} \propto \left(m \Omega_H - \omega_{n\ell m} \right)\alpha^{4\ell + 5} \quad {\rm for} \quad \alpha \ll 1, 
\ea
where $\alpha  \equiv GM\mu/\hbar c$ is the ratio between the Compton wavelength and the size of the BH, $\Omega_H$ is the angular velocity of the BH, and $\omega_{n \ell m}$ is the eigenfrequency of a particular mode denoted by the ``quantum" numbers $\{ n,\, \ell,\,m\}$. An eigenmode grows if $m \Omega_H > \omega_{n\ell m}$, as negative energy flux falls into the black hole horizon. As a result, the BH spins down. Eventually the mode instability saturates when $m \Omega_H \approx \omega_{n\ell m}$, with a long-live cloud rotating around the BH \cite{east2017superradiant,east2017superradiant2}.

The superradiant cloud around a BH could lead to many interesting observational effects. For instance, the cloud can emit monochromic GWs \cite{Yoshino:2013ofa, arvanitaki2015discovering, baryakhtar2017black}, which may be observed from a single newly-formed BH right after binary BH merger \cite{isi2019directed}, or through coherent stacking a set of events \cite{yang2017black}. It may also be detected in all-sky searches \cite{goncharov2018all,pierce2018searching} or as stochastic background by GW detectors such as LIGO and LISA \cite{Brito:2017zvb, Brito:2017wnc}. The evolution of the saturated cloud also has been studied in binary systems \cite{Baumann:2018vus, Berti:2019wnn}. If a superradiant cloud forms around a supermassive BH, it may affect the dynamics of compact objects orbiting around the BH, and leave fingerprints on the waveform of GWs emitted by such objects, providing another approach to search for light bosons with GWs \cite{Ferreira:2017pth, Hannuksela:2018izj, Zhang:2018kib}. 

Previous studies on cloud-induced orbital modulation usually assume fixed cloud density profile based on the mode wave function, so that the gravitational effects of the cloud may be obtained through multipole expansions \cite{Ferreira:2017pth, Hannuksela:2018izj}. This assumption however neglects the cloud density perturbation generated by the gravitational interaction between the cloud and the orbiting object, as demonstrated in \cite{Baumann:2018vus, Berti:2019wnn, Zhang:2018kib}. In particular, it has been shown in \cite{Zhang:2018kib} that, gravitational tidal interaction could deform the cloud in a way that, the backreaction lead to angular momentum transfer from the cloud to the orbiting object. For extreme-mass-ratio inspirals, such angular momentum transfer can be sufficiently strong to compensate the angular momentum loss caused by GW emission, if the object's motion resonantly induces level mixing of the cloud. As a result, the orbit stops decaying and floats at a certain radius, emitting monochromic GWs for a long time until the cloud is depleted. Floating orbits would not be possible if the cloud back reaction is neglected. 

In this work, we further investigate gravitational interactions between superradiant clouds and their surrounding compact objects, focusing on two effects: dynamical friction and backreaction due to level mixing.\footnote{If the object is a stellar mass BH, its motion is also affected by cloud accretion. However, as we will see later, the effect of accretion is negligible comparing to dynamical friction.} When a compact object passes through a medium, the medium usually forms overdensity trail behind the object which exerts gravitational drag on the object, i.e., dynamical friction \cite{Ostriker:1998fa}. While dynamical friction generically occurs for a compact object traveling through extended distribution of matter, the magnitude of the drag force depends on the properties of the component matter, such as the sound speed of the sounding medium. In the case of a scalar field, the effects of dynamical friction depends on the mass of the field \cite{Hui:2016ltb}. We will show that within some mass range the energy loss caused by dynamical friction can be much larger than that caused by GW radiation, with the orbits significantly altered.

On the other hand, level mixing is a specific phenomenon associated with superradiant cloud \cite{Baumann:2018vus, Berti:2019wnn}. Generally speaking, a saturated cloud should be dominated by a particular eigenmode, i.e. the mode that grows fastest, while all modes evolve independently at the linear level. In the presence of an orbiting object, eigenmodes of cloud that evolve independently become coupled under the tidal potential of the orbiting object, and the wave function of the cloud starts oscillating between different modes. If the object is in a quasi-circular orbit, the backreaction of level mixing ``almost" vanishes after averaging over one orbital period. The extra subtlety comes from the fact that the cloud will lose energy to the BH when a decay mode is excited during the Rabi oscillation, i.e. modes with $m \Omega_H < \omega_{n\ell m}$. Because of this dissipation, backreaction on the orbits does not vanish exactly, but is proportional to the decay rate of the decay modes \cite{Zhang:2018kib}. Despite of the small decay rate, the backreaction on the orbits can still be significant, if the mass of the cloud is much larger than that of the orbiting object and the Rabi oscillation is resonantly excited. We shall refer this process as resonant level mixing, as studied in \cite{Zhang:2018kib}. In a separate scenario, i.e., with highly eccentric and hyperbolic orbits, different modes of the cloud usually get dynamically excited after one pericentre passage, which leads to a change in the cloud energy even if mode decay is neglected. In this case, the boson particles redistribute between different levels, and the energy change of the cloud should be proportional to the energy gain/loss while shifting between different modes, i.e., $\Delta E \sim \alpha^2 \mu \Delta c^2$, where $\Delta c$ is the change of the amplitude of subdominant modes. Based on energy conservation, the orbital energy of the object should also change by the same amount, but with a negative sign. We shall refer to this process as dynamical level mixing. Given the equation of motion of the cloud in \eqref{eq:dci}, we expect $\Delta c$ to be proportional to the strength of tidal coupling. The overall effect of dynamical level mixing on the orbits is proportional to the mass ratio between the orbiting object and the BH. It is sub-dominating in extra-mass-ratio systems, but could have interesting impact on comparable-mass binaries.

This paper is organized as follow. In Sec.~\ref{sec:DF} and Sec.~\ref{sec:LM}, we will study the effects of dynamical friction and cloud level mixing respectively. In Sec.~\ref{sec:EMRI}, we will focus on extreme-mass-ratio-inspirals (EMRIs), which are important sources for LISA-like GW detectors. We will investigate the effects of a superradiant cloud on the EMRI waveform as well as on the EMRI rate. In Sec.~\ref{sec:BFR}, we will discuss the effects of a superradiant cloud on stellar mass BH coalescence, and on the formation rate of stellar mass BH binaries, which are the important sources for both space-based and ground-based GW detectors. Sec.~\ref{sec:Dis} is devoted for discussion.

In almost all examples, we consider a scalar field of mass $\mu$ developing a superradiance cloud around a BH of mass $M$. We use the subscription $_{*}$ for quantities associated with the orbiting object, which could be a stellar mass BH or a neutron star. For example, $M_*$ is the mass of the orbiting object, and $\{r_*, \theta_*, \phi_*\}$ are the position of the object in spherical coordinates. We sometimes refer the compact objects as stars, e.g., in Sec.~\ref{sec:EMRI}. But one should keep in mind that they could be both stars and BHs. Unless specified, we assume natural units with $G=c=\hbar=1$.

\section{Dynamical Friction}
\label{sec:DF}

As a compact object moves through the cloud, an overdensity trail forms behind it, which exerts a gravitational drag on the object, i.e. dynamical friction. The friction force can be written as \cite{Chandrasekhar:1943ys, Hui:2016ltb}
\ba\label{eq:FDF}
F_{\rm DF} = \frac{4\pi G^2 M_*^2 \rho}{v^2} C_\Lambda .
\ea
In our case, $v$ is the velocity of the object relative to the wave function of the cloud $\Psi$, $\rho$ is the density of the cloud, and $C_{\Lambda} = C_{\Lambda}\left(\xi,\, k r_{\Lambda}\right)$ with $\xi \equiv G M_* \mu /\hbar v$, $k \equiv \mu v/\hbar$, and $r_{\Lambda}$ representing the smaller quantity between for simplicity the size of the orbit and the size of the cloud. In particular, it is calculated in \cite{Hui:2016ltb} that
\ba
C_\Lambda &\equiv& \frac{e^{\pi \xi} \left|\Gamma\left(1-i\xi\right)\right|^2}{2\xi} \int_0^{2kr_{\Lambda}} dz \left| {\rm F} \left(i\xi,\, 1,\,iz\right)\right|^2  \nonumber \\
&&\times \left(\frac{z}{kr_{\Lambda}} - 2 -\log \frac{z}{2kr_\Lambda}\right),
\ea
where $\Gamma$ is the gamma function and ${\rm F}$ is a confluent hypergeometric function. Following \cite{Hui:2016ltb}, the velocity of the field is defined as
\ba
{\bf v} \equiv \frac{\hbar}{\mu}\, \nabla \Theta,
\ea
where $\Theta$ is the phase of the wave function, i.e., $\Psi \propto e^{-i\left(\omega t - \Theta\right)}$. For fuzzy dark matter discussed in \cite{Hui:2016ltb}, the wave function has no angular dependence and the velocity of the fluid vanishes. In the case of a superradiant cloud, the wave function is rotating with a velocity along the $\phi$-direction,
\ba
{\bf v} = \frac{m\hbar}{r \mu \sin \theta} \, \hat{\phi}.
\ea
Considering a circular orbit, we have
\ba
\frac{\left| {\bf v} \right|}{v_*} \sim \sqrt{\frac{GM}{r \alpha^2}},
\ea
which means the relative velocity is dominated by the orbital velocity if $r > \alpha^{-2}GM$. As most of the cloud mass is in the region outside $\alpha^{-2}GM$, we will approximate the velocity relative the wave function with the orbital velocity for simple. If we consider a stellar mass object orbiting around a supermassive BH, we have
\ba
\xi = \frac{M_*}{M} \frac{\alpha}{v} \ll 1,
\ea
in which case,
\ba
C_\Lambda (kr) = {\rm Cin}(2kr) + \frac{\sin 2kr}{2kr} -1 +{\cal O}(\xi) 
\ea
with ${\rm Cin}(z) \equiv \int_0^z (1-\cos t)dt/t$.

\begin{figure*}[tbp]
\centering 
\includegraphics[width=0.4\textwidth]{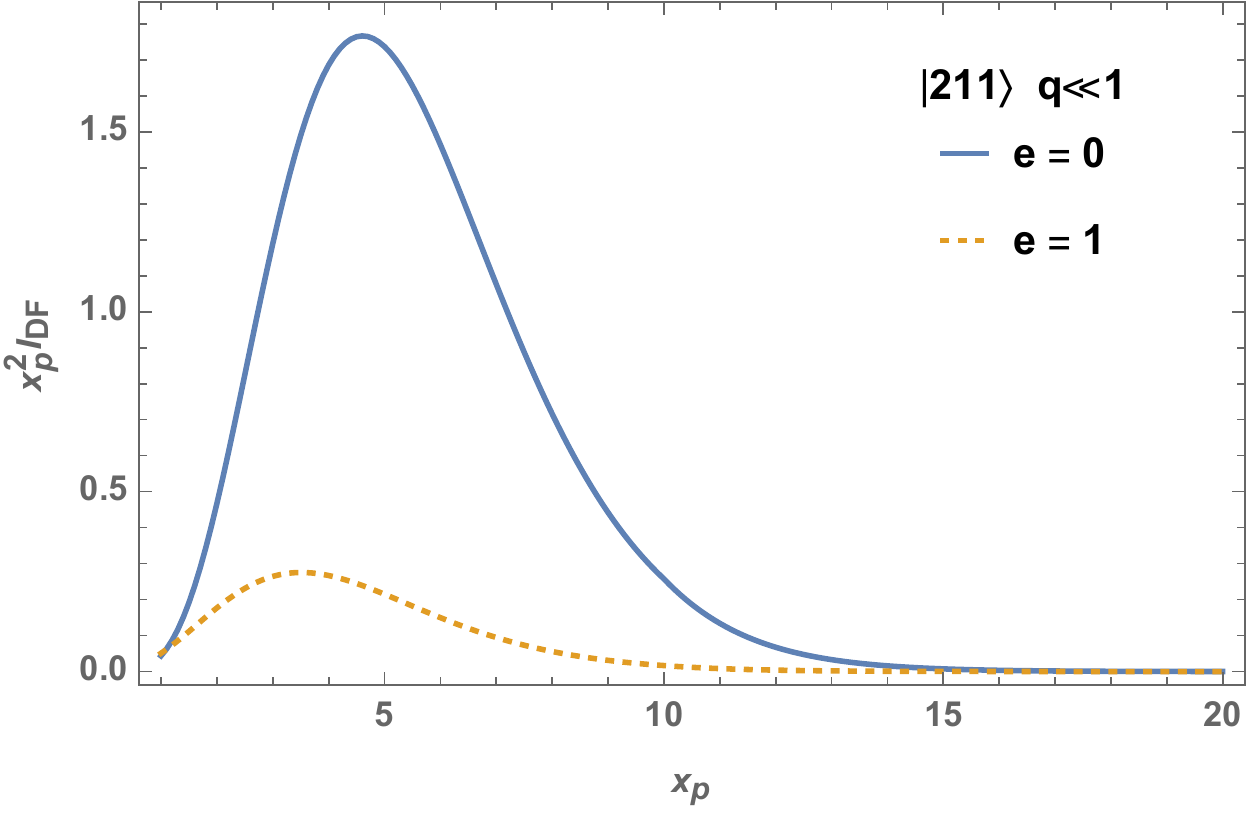}~~~
\includegraphics[width=0.4\textwidth]{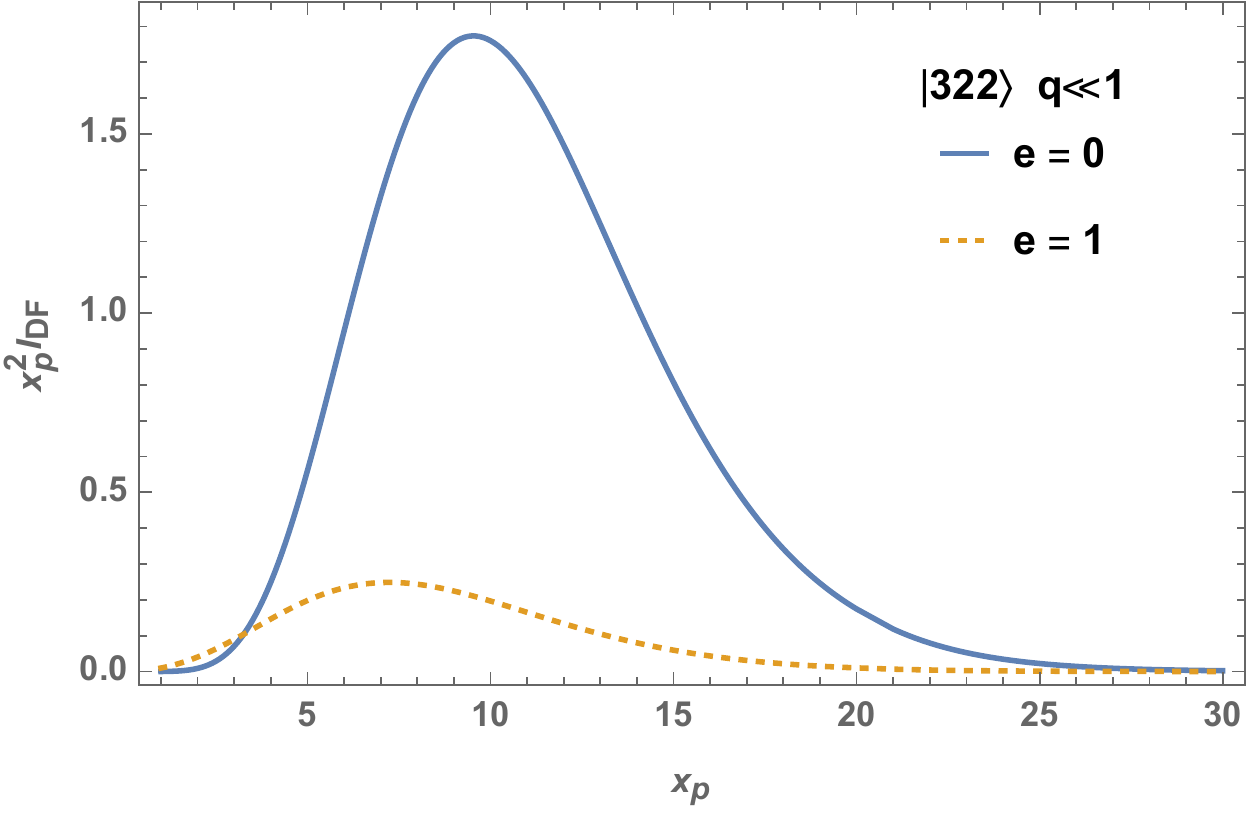}\\ 
\includegraphics[width=0.4\textwidth]{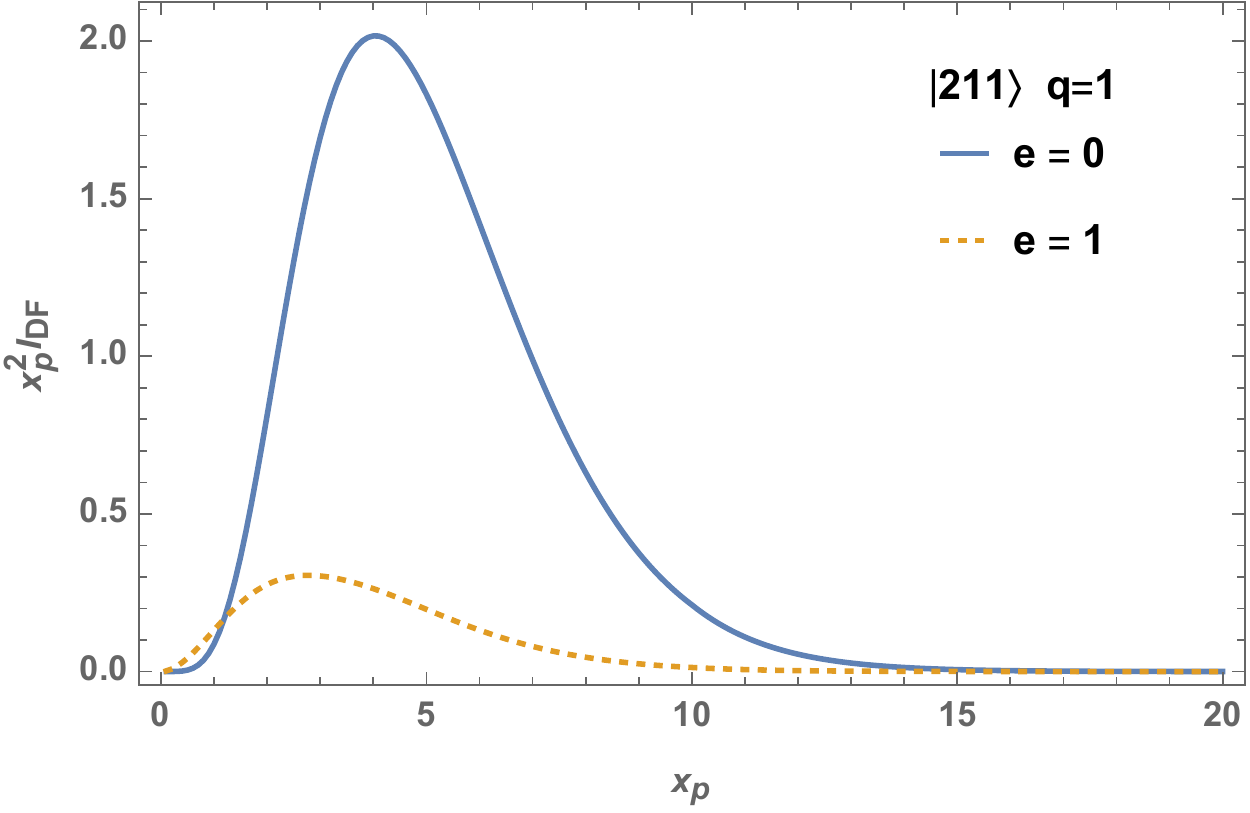}~~~
\includegraphics[width=0.4\textwidth]{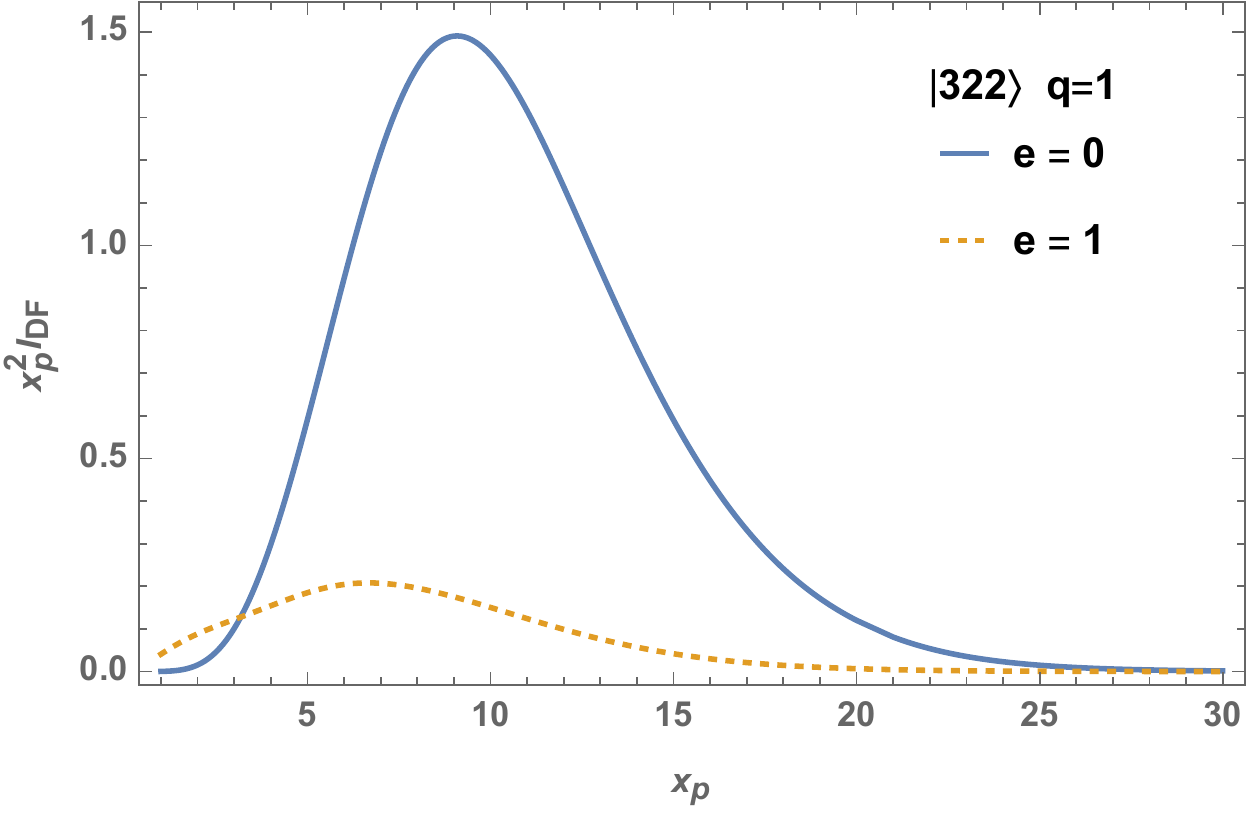} 
\caption{The value of $x_p^2\, \I_{\rm DF}\left[x_p, e, \hat{n} \right]$ with $e = 0$ (blue) and $1$ (orange, dotted), assuming an orbit lies in the equator. $x_p \equiv \alpha^{2} r_p/GM$ with $r_p$ being the periapsis of the orbit. The left panels assume the cloud is saturated at $\ket{211}$ mode, while the right panels assume the cloud is saturated at $\ket{322}$ mode. The upper panels assume $q=M_*/M\ll1$, and the lower panels assume $q=1$.} \label{fig:Is}
\end{figure*}

We assume the orbits are Keplerian at the leading order, in which case we neglect the relativistic corrections as well as the gravitational effects of the cloud. For elliptical orbits, we have
\ba
&&r_*=\frac{a(1-e^2)}{1+e\cos \nu}=a\left(1-e \cos z \right), \\
&&z - e \sin z = \sqrt{\frac{GM}{a^3}}(t-t_0), 
\ea
where $a$ is the semi-major axis of the orbit, $e$ is the eccentricity, and $z$ is the eccentricity anomaly. It is convenient to define the specific energy and the specific angular momentum of the orbit (here after energy and angular momentum):
\ba
\epsilon = \frac{GM}{r} - \frac12 v^2 = \frac{GM}{2a}, \quad J = J_m \sqrt{1-e^2}, 
\ea
where $J_m=\sqrt{GMa}$ is the angular momentum of the circular orbit and is also the maximal angular momentum for a given $\epsilon$.
The energy loss during one orbital period $T$ is
\ba\label{eq:deDF}
\Delta \epsilon_{\rm DF} = \int_0^T \frac{F_{\rm DF}}{M_*}\,  v dt = q \alpha^3 x_p^2\, \I_{\rm DF}\left[x_p, e, \hat{n},\, q \right],
\ea
where $\hat{n}$ is the direction of the orbit angular momentum, and we have defined $q \equiv M_*/M$, $\alpha \equiv GM\mu$ and $x_p \equiv \alpha^{2} r_p/GM$ with $r_p$ being the periapsis of the orbit. $\I_{\rm DF}\left[x_p, e, \hat{n}, q\right]$ is an integral given by
\ba
\I_{\rm DF} &\equiv& \int_0^{2\pi} dz\, \frac{4\pi C_\Lambda}{\left(1-e\right)^2}\, R_{n\ell}^2\left[x_p\left(\frac{1-e\cos z}{1-e}\right)\right] \nonumber \\
&&\times \frac{\left(1-e\cos z\right)^{3/2}}{\left(1+e\cos z\right)^{1/2}}\, Y_{\ell m}^2(\theta_*,\phi_*) ,
\ea
where $Y_{\ell m}$ is the spherical harmonic function and $R_{n\ell}$ is the radial function of the cloud (see App.~\ref{app:SC} for the explicit expression). We also defined $x_{\Lambda} \equiv k r_{\Lambda}$, depending on $x_p$,
\ba
x_{\Lambda} =  
\begin{cases}
\sqrt{x_p}  \frac{\left(1-e\cos z\right)^{3/2}}{\left(1+e\cos z\right)^{1/2}} , & x_p \le x_{97}, \\
\frac{x_{97}}{\sqrt{x_p}} \frac{\left(1-e\cos z\right)^{3/2}}{\left(1+e\cos z\right)^{1/2}} , & x_p > x_{97},
\end{cases}
\ea
where $x_{97}$ represents the size of the cloud and is chosen so that more than $97\%$ of the cloud mass is within $r < x_{97}\,\alpha^{-2}GM$. In Fig.~\ref{fig:Is}, we show $x_p^2\, \I_{\rm DF}\left[x_p, e, \hat{n} \right]$ with $e = 0$ and $1$, assuming the orbit lies in the equator. In the following, we will neglect the $\hat{n}$ dependence for simple. We may expect the result would be changed by an ${\cal O}(1)$ factor after averaged over $\hat{n}$.

As a dissipation effect, dynamical friction accelerates the orbital decay. It would be intuitive to compare the energy loss caused by dynamical friction to that caused by GW radiation, the power of which averaged over one orbital period is \cite{Peters:1963ux}
\ba
P_{\rm GW} = \frac{64\pi}{5} \eta f(e) \left(\frac{r_p}{GM}\right)^{-7/2},
\ea
with $\eta \equiv M_* M/(M_*+ M)^2$ being the dimensionless reduced mass and
\ba
f(e) = \frac{1+(73/24)e^2+(37/96)e^4}{(1+e)^{7/2}}.
\ea
Together with Eq.~\eqref{eq:deDF}, we find
\ba\label{eq:DFGWratio}
\frac{P_{\rm DF}}{P_{\rm GW}}= \frac{5}{64\pi}\frac{\alpha^{-4}}{f(e)} x_p^{11/2} \I_{\rm DF} \left[x_p, e\right],
\ea
which means for orbits with radius comparable to the size of the cloud, the energy loss could be enhanced by roughly a factor of $\alpha^{-4}$ if $\alpha \ll 1$.

If the compact object is a stellar mass BH, it will continuously absorb the cloud, resulting in a force of $F_{\rm Ab}= \sigma_{\rm Ab} \rho v^2$, where $\sigma_{\rm Ab}$ is the absorption cross section and $v$ is the relative velocity between the BH and the cloud. The absorption cross section of a massive scalar field with $M_* \mu \ll 1$ has been calculated in \cite{Benone:2014qaa, Benone:2017xmg, Unruh:1976fm},
\ba
\sigma_{\rm Ab} \simeq \frac{32 \pi^2 G^2 M_*^2 \alpha }{v^2}. 
\ea
Together with Eq.~\eqref{eq:FDF}, we find that
\ba
\frac{F_{\rm Ab}}{F_{\rm DF}} \simeq \frac{8\pi \alpha v^2}{C_\Lambda} \ll 1,
\ea
which means that the force caused by absorption is negligible comparing to dynamical friction.

\section{Level mixing}
\label{sec:LM}

In the presence of a compact object, the gravitational potential of the object will introduce couplings between the modes of the cloud, which mix modes with different energy levels. The effects of level mixing on the superradiant cloud have been studied in \cite{Baumann:2018vus, Berti:2019wnn}, assuming the object is in a quasi-circular orbit or an elliptical orbit. In particular, in \cite{Zhang:2018kib} we have shown that a perturbed cloud backreacts on the orbit as well. In the adiabatic limit, the backreaction on the quasi-circular orbits is proportional to the decay rate of the decay modes, because the energy/angular momentum transfer between the cloud and the object averaged over one orbit is effectively zero if the modes do not decay. However, it is no longer true for hyperbolic orbits or highly eccentric orbits, in which case the backreaction could be significant. When the object approaches the BH, the dominated mode of the cloud will mix with the other modes. After one passage, the wave function of the cloud will be redistributed to each mode, resulting in a change in the cloud's energy, and hence in the orbital energy as well. In this section, we will investigate the backreaction of level mixing on orbits with eccentricity $e \sim 1$. We will see that counter-rotating orbits always deposit energy to the cloud after one passage, while a co-rotating orbit may gain energy from the cloud if the periapsis is close to the radius of the cloud, and will lose energy to the cloud if the periapsis is relatively far from the cloud. We will first discuss the gravitational effects of the orbiting object on the cloud, and then estimate the energy change of orbits after one passage. A similar calculation for tidal excitation of star oscillations for arbitrary eccentricity orbits can be found in \cite{press1977formation,yang2018evolution,yang2019inspiraling}.

Throughout this paper we assume that the cloud size is much larger than the BH size, so that we can adopt a Newtonian approximation for the cloud. For full treatment in the relativistic setting, one can read \cite{Zhang:2018kib} based on techniques developed in \cite{mark2015quasinormal,yang2015turbulent,yang2016plasma}. Let us write the wave function of the cloud as $\ket{\psi} =\sum_i c_i(t) \ket{\psi_i}$, where subscript $i$ is a shorthand for $n\ell m$, and $\ket{\psi_i}$ is the wave function of the eigenmode denoted by $i$. Initially, the cloud is dominated by the saturated mode, say the mode with $i=s$, and hence we have $c_s\simeq 1$, while all the other $c_i \simeq 0$. In the presence of an object, we have 
\ba\label{eq:dci}
i \frac{d c_i}{dt} = \sum_{j} \bra{\psi_i}V_*\ket{\psi_j} c_j,
\ea
where $V_*$ is the tidal perturbation generated by the approaching object. See App.~\ref{app:tide} for more details. The inner product is defined as an integral weighted by wave functions. 

It is important to note a few issues regarding $\bra{\psi_i}V_*\ket{\psi_j}$. First of all, the monopole and the dipole pieces of the tidal potential do not contribute to mode coupling if the object is far away from the cloud \cite{Baumann:2018vus}. It is because the monopole does not lead to a shift in the energy level, and the dipole is fictitious by virtue of the equivalence principle (see the appendix in \cite{Baumann:2018vus} for explicit calculations). Second, $\bra{\psi_i}V_*\ket{\psi_j}$ is Hermitian,\footnote{Here we have neglected the mode decay caused by the non-zero but tiny imaginary eigenfrequency.} which implies $\sum_i \left|c_i\right|^2 = 1$. Moreover, the tidal coupling is proportional to the perturber's mass, so that $\bra{\psi_i}V_*\ket{\psi_j} \propto q$. The conservation of wave function implies that the energy change of the cloud is of order 
\ba
\Delta\e \sim \Delta \e_{ij}\, \Delta \left|c_i\right|^2 \sim \frac{\alpha^2 M_c}{2n^2} \left| q c_{s}\Delta t \right|^2 \propto \frac{\alpha^3 q^2 M}{2n^2}, 
\ea
which is suppressed by the factor $q^2$. Here $\Delta \e_{ij} \sim \alpha^2 M_c/2n^2$ is the energy difference between mode $i$ and mode $j$, and $M_c \simeq \alpha M$ \cite{Brito:2017zvb} is the mass of the cloud. Third, in computing Eq.~\eqref{eq:dci} (or equivalently Eq.~\eqref{eq:aij}), $\ell_*$ should be summed over all $\ell_*$ that subject to the selection rules $\left|\ell_j-\ell_i\right| \le \ell_* \le \ell_j+\ell_i$ \cite{Baumann:2018vus}. The coupling strength $\bra{\psi_i}V_*\ket{\psi_j}$, however, is typically suppressed by $1/r_*^{\ell_*+1}$ when the object is outside of the cloud. Therefore, Eq.~\eqref{eq:dci} is dominated by the modes near the saturated mode in the $\ell$ space. Similarly, although the energy difference between levels of different $n$ is roughly of the same order, i.e. $\Delta \omega_{ij} \sim \alpha^2 \mu$, so that all these modes should be excited during the pericentre passage, the coupling strength decreases if the difference between $n$ of two modes increases. Therefore, it is sufficient in practice to only consider modes that are near the saturated mode for different $n$'s. 

In the following, we approximate hyperbolic orbits and highly eccentric orbits with parabolical orbits, 
\ba
&&r_*=r_p(1+z^2), \quad z=\tan\frac{\nu}{2}, \\
&&t = \sqrt{\frac{2r_p^3}{G \left(M+M_*\right)}}\left(z+\frac13 z^3\right),
\ea
where $\nu$ is the true anomaly. This is a good approximation, as the orbits that are of interest in a capture process usually have $\left|1-e\right| \ll 1$. Under this parametrization, it would be convenient to solve $c_i$ in terms of $z$ instead of $t$. In this case, Eqs.~\eqref{eq:dci} become
\ba\label{eq:dci2}
i \frac{dc_i}{dz} = A_{ij} c_j
\ea
with
\ba\label{eq:aij}
A_{ij} = i q \sqrt{2 x_p^3} \sum_{\ell_*,m_*}\frac{4\pi}{\ell_*+1} \I_{\Omega} \I_{ij}\left(x_*\right) Y_{\ell_* m_*}(\theta_*, 0) \nonumber \\
\times \exp \left[i\left(\omega_{ij} \sqrt{2x_p^3}\left(z+z^3/3\right) \mp m_* \phi_* \right)\right],
\ea
where $\omega_{ij} = \left(\omega_i-\omega_j\right)/\mu\alpha^2$, and the $-$ sign in front of $m_*\phi_*$ is associated with co-rotating orbits, while the $+$ sign is for counter-rotating orbits. For $\ell_* \ge 2$, we have
\ba
\I_{ij}&=&\int_0^{x_*} dx \, x^2 \frac{x^{\ell_*}}{x_*^{\ell_*+1}} R_{n_i \ell_i}(x) R_{n_j \ell_j}(x) \nonumber \\
&&+\int_{x_*}^{\infty} dx \, x^2 \frac{x_*^{\ell_*}}{x^{\ell_*+1}} R_{n_i \ell_i}(x) R_{n_j \ell_j}(x),
\ea
and for $\ell_*= 1$ and $0$, we have
\ba
\I_{ij} = \int_{x_*}^{\infty} dx \, x_* \left(1- \frac{x^3}{x_*^3}\right) R_{n_i \ell_i}(x) R_{n_j \ell_j}(x)
\ea
and 
\ba
\I_{ij} = \int_{x_*}^{\infty} dx \, x R_{n_i \ell_i}(x) R_{n_j \ell_j}(x) 
\ea 
respectively. Given the initial conditions $c_s=1$ while other $c_i = 0$ as $z \rightarrow -\infty$, we can solve $c_i$ at $z \rightarrow +\infty$ numerically. The energy change of the cloud after one passage is given by
\ba
\Delta \e = \frac{M_c}{\mu}\sum_i \omega_i \left( \left|c_i^{+}\right|^2 - \left|c_i^{-}\right|^2\right),
\ea
where the superscriptions $+$ and $-$ denote $z$ approaches $+\infty$ and $-\infty$ respectively. 

\begin{figure}[tbp]
\centering 
\includegraphics[width=0.45\textwidth]{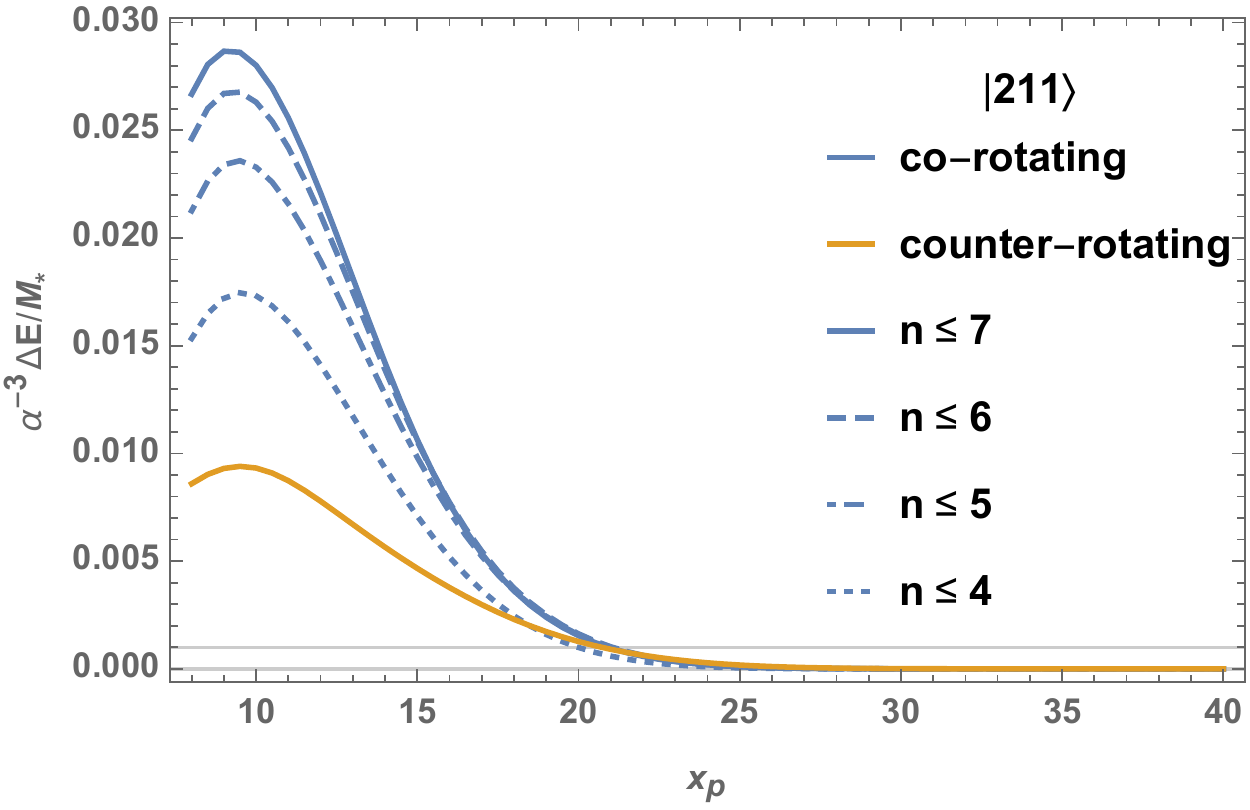} 
\includegraphics[width=0.45\textwidth]{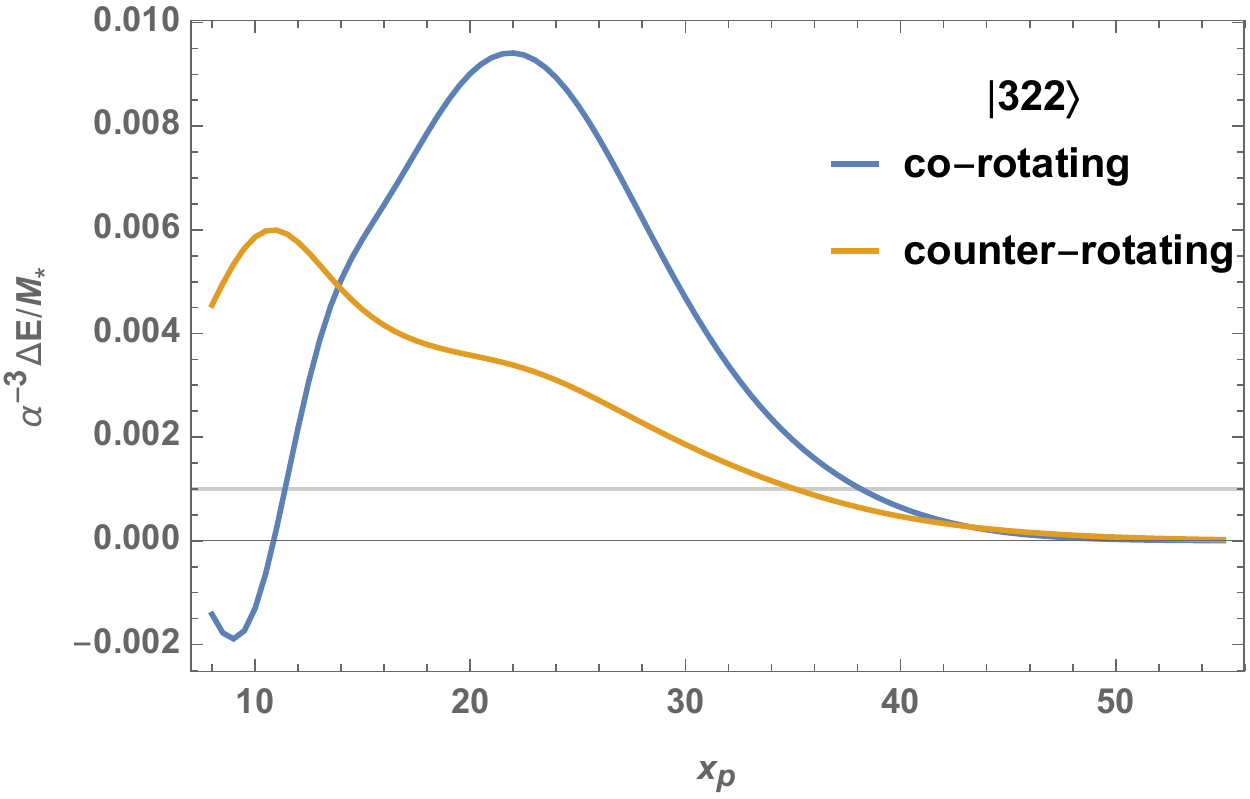} 
\caption{The energy change of the cloud after one passage. We consider parabolic orbits with periapsis $r_p = x_p\, \alpha^{-2}GM$ and $\theta_*=\pi/2$, and choose the mass ratio $q = 1$. The upper panel shows the case with a cloud initially saturates at mode $\ket{211}$, while the lower panel shows the case with a cloud initially saturates at mode $\ket{322}$. We show both co-rotating orbits (in blue) and counter-rotating orbits (in orange) considering all modes with $n \le 7$ (solid lines). We also show the results with different number of modes taken into account ($n \le 4,\,5,\,6$ with dashed, dash-dotted, and dotted lines). We can see that the result converges as $n$ increases. We mark $\I = 0.001$ defined in Eq.~\eqref{eq:Idef} with grey horizontal lines in the plots.} \label{fig:de}
\end{figure}

In a simple scenario, we consider parabolic orbits that lie in the equator, in which case $\theta_*=\pi/2$ and $\phi_*=2 \arctan z$. Therefore the orbit is determined given the parameter $x_p$, i.e. the dimensionless periapsis. We assume a mass ratio of $q=1$ and a cloud initially saturated at $\ket{211}$ or $\ket{322}$, and solve Eqs.~\eqref{eq:dci2} numerically taking into account all the modes with $n \le 7$. The energy change of the cloud is plotted in Fig.~\ref{fig:de}. Inclusion of higher $n$ modes is computational expensive. On the other hand, we are more interested in the range of $x_p$ that are capable to capture the object. Including modes with $n > 7$ does not change this range very much as the energy change is exponentially suppressed at large $x_p$, and is not necessary for the discussion in this paper.

The net change of the cloud energy depends on the modes that are efficiently excited. In other words, the cloud will lose energy if the excitation is dominated by modes with energy lower than the initially saturated mode, and vice versa. On the other hand, according to Eq.~\eqref{eq:dci}, a mode $j$ responds to the tidal force efficiently if $(\omega_{s}-\omega_j)/(m_s-m_j)$ is comparable to the typical orbital frequency.\footnote{Recall that $\bra{\psi_i}V_*\ket{\psi_j } \neq 0$ only if $m_*=m_i-m_j$ \cite{Baumann:2018vus}.} In order to take into account both co-rotating and counter-rotating orbits, we say the orbital frequency is positive if the orbit is co-rotating and is negative if the orbit is counter-rotating. Therefore, a counter-rotating orbit always rises the energy of the cloud, as there is no lower energy level with $m_j > m_{s}$. For co-rotating orbits, the cloud can lose energy, if the tidal force excites the lower energy modes with $m_j < m_{s}$. As the energy difference between the saturated mode and a lower energy mode is usually larger than that between the saturated mode and a higher energy mode, we may expect the lower energy modes will get efficiently excited when the typical orbital frequency is larger or equivalently when $x_p$ is small.

As shown in the upper panel of Fig.~\ref{fig:de}, if the cloud initially saturates at mode $\ket{211}$, a co-rotating orbit can only make it couple to higher energy modes,\footnote{Mode $\ket{211}$ can couple to lower energy modes with $n=1$ only through the monopole and dipole of $V_*$, which are effectively zero when the star is far away from the cloud.} therefore the object can only lose energy to the cloud. If the cloud initially saturates at $\ket{322}$ mode, a co-rotating orbit can couple it to lower energy modes, such as modes with $n=1$. As $x_p$ decreases, the excitation is dominated by the $n=1$ modes, which explains the valley in the lower panel of Fig.~\ref{fig:de}. Here we assume modes with same $n$ have the same energy. In principle, they have different energy due to the higher order corrections to the BH potential. However, the energy difference between modes with same $n$ is further suppressed by a factor of $\alpha^2$, which is negligible. Depending on the periapsis of the orbit, the modes of the cloud get exited to different extent, resulting a different amount of exchanging in total energy.

\section{Effects on EMRIs}
\label{sec:EMRI}

In this section, we discuss the observational effects of a superradiant cloud on EMRIs. For quasi-circular orbits, the observational effects of level mixing has been studied in \cite{Zhang:2018kib} (also see App.~\ref{app:MD} for further discussion). For hyperbolic and highly eccentric orbits, which are of most interest in the estimation of EMRI rate, the effects of level mixing is suppressed by the mass ratio $q$ to be negligible. Therefore, we will mainly focus on dynamical friction in the following discussion. 

\subsection{Inspiral waveform}

\begin{figure}[tbp]
\centering 
\includegraphics[width=0.45\textwidth]{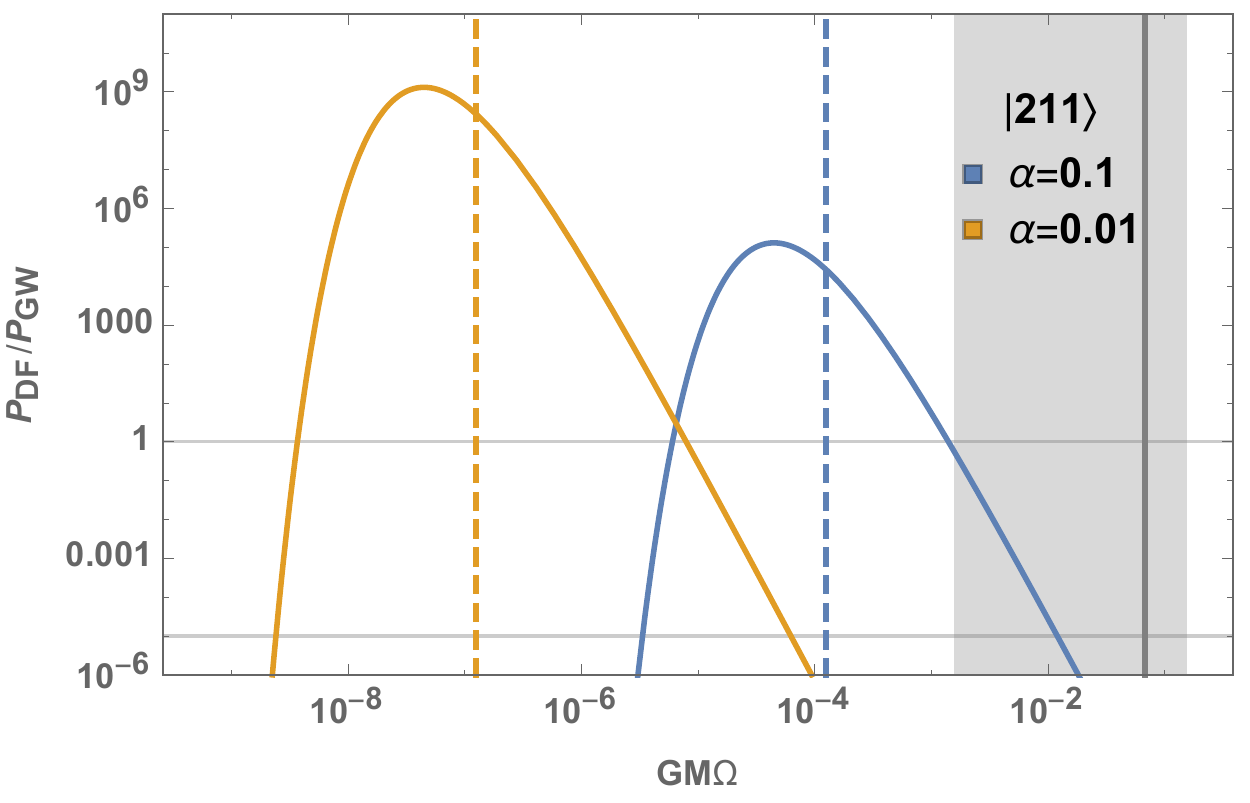} 
\includegraphics[width=0.45\textwidth]{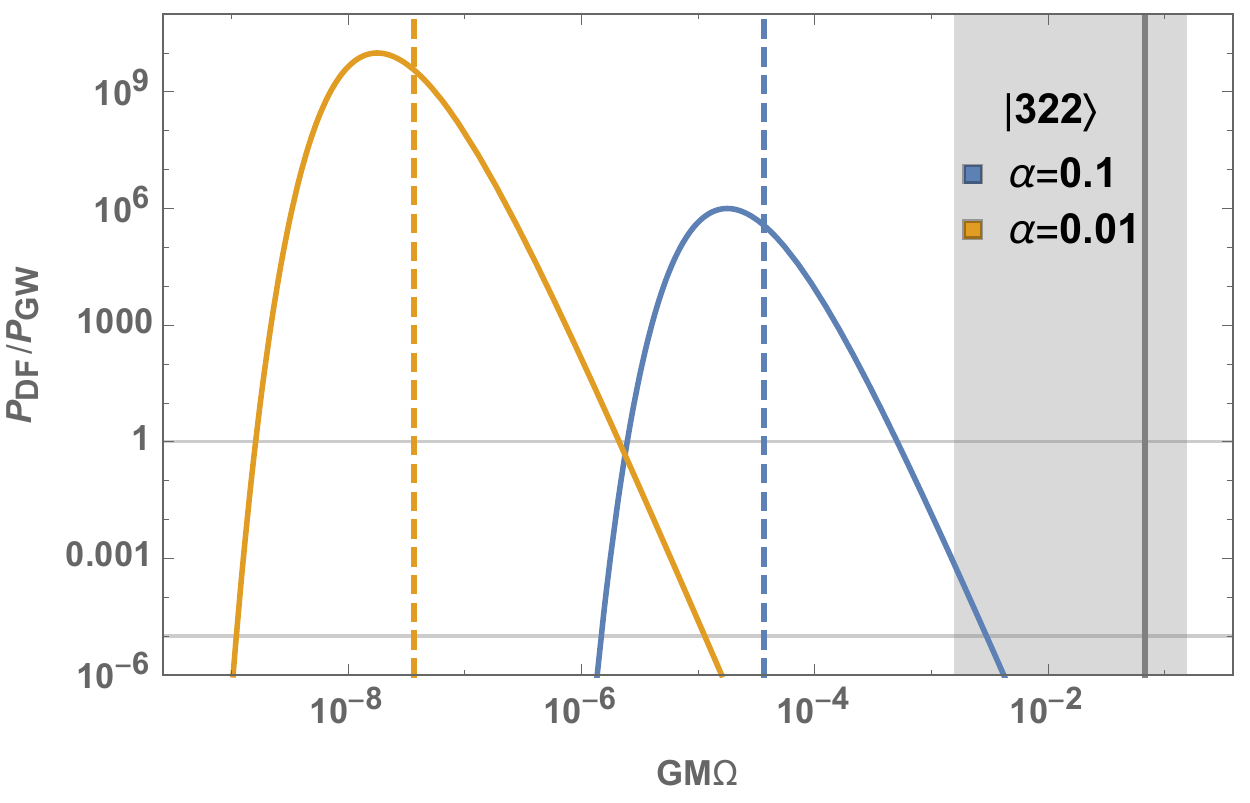} 
\caption{The ratio between energy loss rate caused by dynamical friction and that caused by GW radiation. In the plot, we consider a cloud saturated at $\ket{211}$ mode (upper panel) or $\ket{322}$ mode (lower panel). The horizontal axis shows the orbital frequency, scaled by the mass of the black hole. The blue/orange dashed line shows the orbital frequency corresponding to a radius of $n^2\alpha^{-2}GM$, which locates the density peak of the cloud. The shaded region shows the LISA frequency band ($0.1$Hz - $0.001$Hz), assuming a $10^5 M_{\odot}$ BH. The gray vertical line denotes the innermost stable circular orbit. The observation band shifts toward the left for smaller BHs.} \label{fig:DF}
\end{figure}

As shown in Sec.~\ref{sec:DF}, dynamical friction accelerates orbit decay and introduces phase shift in the EMRI waveform. In fact, dynamical friction can be much more efficient in draining orbital energy as compared to GW radiation, when the orbiting object immerses in the cloud. However, LISA usually starts seeing EMRIs when the pericentre distance is at least within 20 $GM$ away from the BH, while the density peak of the cloud is much further sway. Therefore, it is unlikely to observe the effect of dynamical friction in the waveform, unless the central BH is less massive than $10^5 M_\odot$.

For example, let consider a circular obit lying in the equator, and compare the energy loss power caused by dynamical friction $P_{\rm DF}$ to that caused by GW radiation \cite{Peters:1963ux}
\ba
P_{\rm GW} \simeq -\frac{32}{5}GM^2 q^2 r_*^4 \Omega^6,
\ea
where we have used $\eta \simeq q$ for $M \gg M_*$. We plot the dependence of $P_{\rm DF}/P_{\rm GW}$ on the orbital frequency $\Omega$ in Fig.~\ref{fig:DF}. Note that the ratio $P_{\rm DF}/P_{\rm GW}$ does not depend on $q$. For a LISA-like GW detector, the effects of dynamical frication may be detected if $P_{\rm DF}/P_{\rm GW} \ge 10^{-5}$ in the observation band, as the number of cycles staying in band could be order $10^5$. Taking $M=10^5 M_\odot$ as a benchmark, we find that this requires $\alpha > 0.1$. However, in such case the cloud might be depleted by the tidal perturbation of the earlier falling object as suggested in \cite{Baumann:2018vus, Berti:2019wnn} or the supperradiant spin-down, unless the BH angular momentum is fed by accretion efficiently. The requirement of $\alpha > 0.1$ can be relaxed for clouds around intermediate mass BHs.

\subsection{EMRI rate}

Besides the waveform study for individual sources, another important message we may receive from a population of future EMRI observations is the EMRI merger rate, which encodes information of galactic-centre stellar distributions \cite{berry2019unique}. The theoretical EMRI rate can be estimated by virtue of loss cone dynamics, which will be briefly summarized as follows (also see \cite{Merritt:2013cqg} for a review on loss cone dynamics). 

Assuming spherical symmetry, orbits around the central BH can be specified by two parameters, i.e., energy $\epsilon$ and angular momentum $J$. Taking into account gravitational encounters, the distribution of the orbits around a supermassive BH can be obtained by solving the Fokker-Planck equation. Given the orbit distribution, the rate of the stars\footnote{Here we refer all compact objects as stars for simplicity, but keep in mind that they can be neutron stars or stellar mass BHs.} falling into the BH can be found by calculating the flux of the stars into the loss cone, a regime defined in the two-parameter phase space and in which stars fall into the supermassive BH in one orbital time. In particular, orbits with higher magnitude of energy have short periods,\footnote{Note that in our notation of energy, orbits of smaller semi-major axis have higher energy.} and stars in such orbits hardly penetrate beyond the loss-cone boundary before they are consumed by the BH. This defines the empty-loss-cone, or the diffusive regime of the phase space. On the other hand, it is possible for a star in a low energy orbit to diffuse across the loss cone by gravitational encounters during a single orbital period. This defines the full-loss-cone, or pinhole regime of the phase space. It has been shown that both regimes contribute to the flux of stars falling into the BH. 

However, a star falling into the BH does not necessarily lead to an EMRI. To secure a successful EMRI, the star should avoid scattering during the orbital decay. That means $t_0 < t_J$, where $t_0$ is the lifetime of the orbit, and the $t_J$ is the angular momentum relaxation time. The criterion $t_0 < t_J$ indicates that the full loss cone regime barely contributes to the EMRIs. Stars in such regime are scattered multiple times each orbit. The probability of the same star falling into the BH in the corresponding orbit lifetime, even in the presence of a large number of gravitational encounters, is effectively zero. The criterion $t_0 < t_J$ defines a critical radius $a_c$ (or equivalently, a critical energy $\e_c$), such that stars starting inspiral from an orbits with $a < a_c$ are able to fall into the central BH with high probability. Therefore, the inspiral rate can be estimated as
\ba\label{eq:EMRIInter}
\Gamma_{\rm EMRI} \simeq \int_0^{a_c} da\, {\cal F}(a),
\ea
where ${\cal F}(a)$ is the flux of stars into the loss cone per radius interval. Contributions to EMRIs are mainly from the diffusive regime, the orbit distribution in which can be well approximated by the steady state distribution obtained from the one-dimensional (angular momentum-dependent) Fokker-Planck equation. In this case the star flux into the loss cone is
\ba
{\cal F}(a) = \frac{N\left(a\right)}{ \ln \left(J_m/J_{lc}\right)t_r(a)},
\ea
where $N(a)$ is the number of stars within $a$, $J_m(a) =\sqrt{GMa}$ is the maximal (circular orbit) angular momentum for a specific energy $\e$, and $t_r(a)$ is the angular momentum relaxation time at radius $a$. In the following, we will first estimate the inspiral rate considering only GW radiation. We will mostly follow the estimation in \cite{Hopman:2005vr}. After that we will discuss how the inspiral rate is changed in the presence of a superradiant cloud. 

Without the cloud, the energy dissipation is caused by GW radiation (we neglect other dissipation effects for simplicity), and the lifetime of the orbit is 
\ba
t_0^{\rm GW} = \int_{\e_0}^{\infty} \frac{d\e}{d\e/dt} \simeq \frac{2\pi \sqrt{GM a}}{\Delta \e_{\rm GW}},
\ea
where $\Delta \e_{\rm GW}$ is the energy loss caused by GW radiation in one orbital period. GWs also carry away angular momentum. Generally, the changes of $J$ during inspiral is dominated by two-body scattering, and $\Delta J_{\rm GW}$ can be neglected until $a$ becomes very small. Given the stellar velocity dispersion of the host bulge $\sigma$, we can define the radius of influence of the supermassive BH $r_h \equiv GM/\sigma^2$. It is convenient to refer the relaxation time to the relaxation time at the BH radius of influence,
\ba
t_h = A_p\, q^{-2} \frac{T(r_h)}{N_h \log \Lambda_1},
\ea
where $T(r_h)$ is the orbital period corresponding to $r_h$, $N_h$ is the number of stars within $r_h$, and $\Lambda_1 = q^{-1} \left(2GM/r_h\right)^{1/4}$. In terms of $t_h$, the relaxation time at any radius $a$ is given by
\ba
t_r(a) = t_h \left(\frac{a}{r_h}\right)^p,
\ea
and the angular momentum relaxation time is
\ba\label{eq:tJ}
t_J = \left[\frac{J}{J_m(a)}\right]^2 \left(\frac{a}{r_h}\right)^p t_h.
\ea
Simulations indicate that $0 \le p \le 0.25$ \cite{Freitag:2002mj,Baumgardt:2004zq,Baumgardt:2004zu,Preto:2004kd}, which means $t_r$ is roughly independent of radius $a$. In particular, we have $A_p \simeq 0.2$ for $p=0$ \cite{Alexander:2003nc}. Therefore, the criterion for a successful EMRI becomes
\ba\label{eq:cri}
\frac{t^{\rm GW}_0}{t_J} = \left(\frac{a}{r_h}\right)^{\frac{3}{2}-p} \zeta_{\rm GW}^{\frac32-p}(J)< 1,
\ea
where we have used the fact that $r_p/GM = 8 (J/J_{lc})^2$ for orbits with $e \simeq 1$, and have defined
\ba
\zeta_{GW} (J) \equiv \left[\left(\frac{J}{J_{lc}}\right)^{-5} \left(\frac{85\sqrt{GM} t_h}{3\times 2^{10} r_h^{3/2}}\right) \right]^{\frac{2}{3-2p}}.
\ea
Here $J_{lc}= \sqrt{2 r_{lc}^2 (\epsilon - GM/r_{lc})} \simeq \sqrt{2GM r_{lc}}$ is the loss-cone angular momentum corresponding to a loss cone radius $r_{lc}$, which is about a few $GM$. 
As we mentioned $J$ does not change too much for large $a$, and therefore $J \sim J_{lc}$. As a benchmark, we take $p=0$ and $M = 10^6 M_\odot$. In this case, we have $\zeta_{\rm GW} \sim 0.016$ and $a^{\rm GW}_c = \zeta_{\rm GW} r_h \ll r_h$. That means orbits leading to successful EMRIs usually have radius much smaller than the radius of influence. Integrating Eq.~\eqref{eq:EMRIInter} up to the critical radius gives
\ba
\Gamma_{\rm EMRI} \sim \frac{N_h}{t_h \log \left[J_m(a^{\rm GW}_c)/J_{lc}\right]} \left(\frac{a^{\rm GW}_c}{r_h}\right)^{3/2-2p}.
\ea

Now let us discuss how dynamical friction affects the inspiral rate. We will show that dynamical friction does not affect the orbit with $r_p \simeq r_{lc}$. For $\alpha \ll 1$, the cloud is expected to be far from the BH. The orbits with $r_p \simeq r_{lc} \sim 10 GM$ are affected by the cloud only if $e \simeq 1$. Nevertheless, for these orbits the energy loss is dominated by GW radiation, as one can check $\Delta \e_{\rm DF}/\Delta \e_{\rm GW} \sim 0.06\,\alpha^7 \I_{\rm DF} \ll 1$. On the other hand, dynamical friction dominates GW dissipation if the pericentre distance is comparable to the size of the cloud: $r_p \sim \alpha^{-2}GM$. In this case, orbits with $r_p \sim \alpha^{-2}GM$ are first circularized by the cloud in a timescale $t^{\rm DF}_0 \ll t^{\rm GW}_0$ and then continuously inspiraling into the MBH due to GW radiation. For a given orbit with semi-major axis $a$, we have 
\ba
t_0^{\rm DF} \simeq \frac{2\pi \sqrt{GM a}}{\Delta \e_{\rm DF}}.
\ea
We still need to impose the condition that $r_p < (J/J_{cl})^{-10/(3-4p)} a^{\rm GW}_c$ according to Eq.~\eqref{eq:cri}. That is the orbital life time is shorter than the angular momentum relaxation time, otherwise the orbit cannot complete the inspiral stage even after the circularization. Here we keep the $J$-dependence in Eq.~\eqref{eq:tJ} as $J$ is no longer $J_{lc}$ as in the case of GW dissipation. By doing this, we take into account the contribution from orbits with small eccentricity. To summarize, in the presence of the cloud, an orbit with radius $a$ could complete inspiral if it can be circularized down to a radius of $\alpha^{-2}GM$ in a timescale $t_0^{\rm DF} \ll t_J$, and if $r_p < (J/J_{lc})^{-10/(3-4p)} a^{\rm GW}_c$. Assuming $x_p \simeq 1$, we can find that the former condition defines a critical radius
\ba
a^{\rm DF}_c &=& \left(\frac{J}{J_{lc}}\right)^{\frac{-10}{3-2p}} \left(\frac{\Delta \e_{\rm DF}}{\Delta \e_{\rm GW}}\right)^{\frac{2}{3-2p}}\, a^{\rm GW}_c \\
&\simeq& \left(\frac{3\times 2^{10}\I_{\rm DF}}{85\pi} \alpha\right)^{\frac{2}{3-2p}} a^{\rm GW}_{c}.
\ea
For $M=10^6 M_\odot$ and $p=0$, we have $ a^{\rm DF}_c \simeq 0.6 \alpha^{2/3} a^{\rm GW}_{c}$, which is usually smaller than $a^{\rm GW}_{c}$. Therefore, the EMRI rate is still decided by GW dissipation, and dynamical friction caused by the superradiant cloud should not alter the EMRI rate significantly. This result is expected, because usually the EMRI rate is not determined by capture mechanism, but by how efficient the loss cone gets repopulated, where the later is still determined by gravitational scattering even in the presence of the superradiant cloud.

However, a superradiant cloud can affect the eccentricity distribution of EMRIs. During orbital decay, the evolution of the eccentricity $e$ in terms of the semi-major axis $a$ is given by
\ba
\frac{\d e}{\d a} = -\frac{\sqrt{1-e^2}}{e}\frac{\sqrt{GM}}{2 a^{5/2}} \left< \frac{\d L}{\d t}\right> \left< \frac{\d E}{\d t}\right>^{-1} + \frac{1-e^2}{2 e a},
\ea
where $\left< \frac{\d L}{\d t}\right>$ and $\left< \frac{\d E}{\d t}\right>$ are the angular momentum and the energy loss rate averaged over one orbital period. If the orbital decay is dominated by GW radiation, we have
\ba
\frac{\d e}{\d a} = \frac12 \frac{1-e^2}{e a} \left(\frac{208 e^2 + 37 e^4}{96 + 292 e^2 +37 e^4}\right).
\ea
If the orbital decay is dominated by a friction $F$, we have
\ba
\left< \frac{\d L}{\d t}\right> = \frac{1}{2\pi} \int \frac{r^{3/2}F}{a\sqrt{2a-r}} d\theta
\ea
and
\ba
\left< \frac{\d E}{\d t}\right> = \frac{1}{2\pi} \int  \frac{r^2F}{a^2\sqrt{1-e^2}} \sqrt{\frac{2GM}{r}-\frac{GM}{a}} d\theta,
\ea
with $r=a(1-e^2)/(1+e \cos \theta)$, which generally results in a different evolution of eccentricity as it is dominated by GW radiation.

\section{Effects on stellar mass binary black hole coalescence}
\label{sec:BFR}

Stellar mass BH binary is one of the main sources for future spaced-based and ground-based GW detectors, and some of them will be detected by both types of detectors to allow multiband analysis \cite{sesana2016prospects,vitale2016multiband,wong2018expanding,cutler2019we}. In this section, we would like to investigate the effects of a superradiant cloud on inspiral wavefom of stellar mass BH binaries, as well as on the binary formation rate. 

\subsection{Inspiral wavefom}

\begin{figure}[tbp]
\centering 
\includegraphics[width=0.45\textwidth]{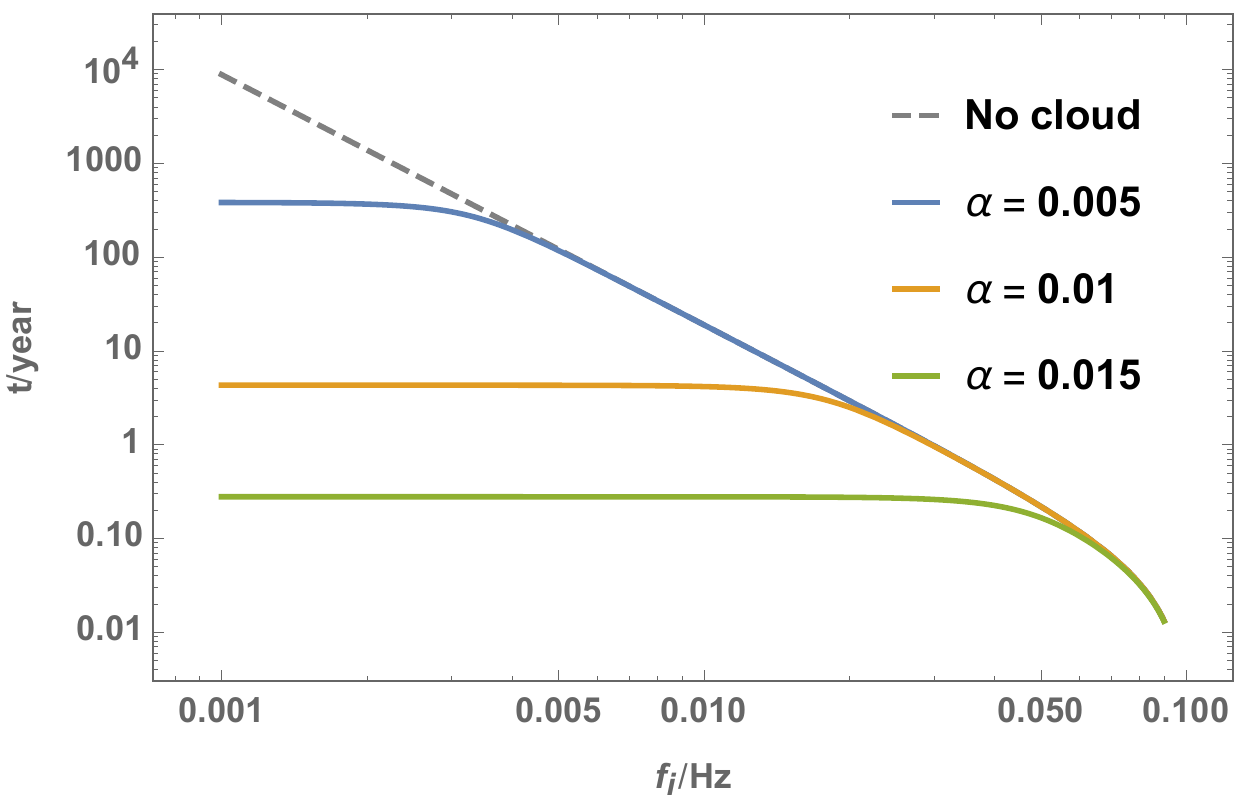} 
\includegraphics[width=0.45\textwidth]{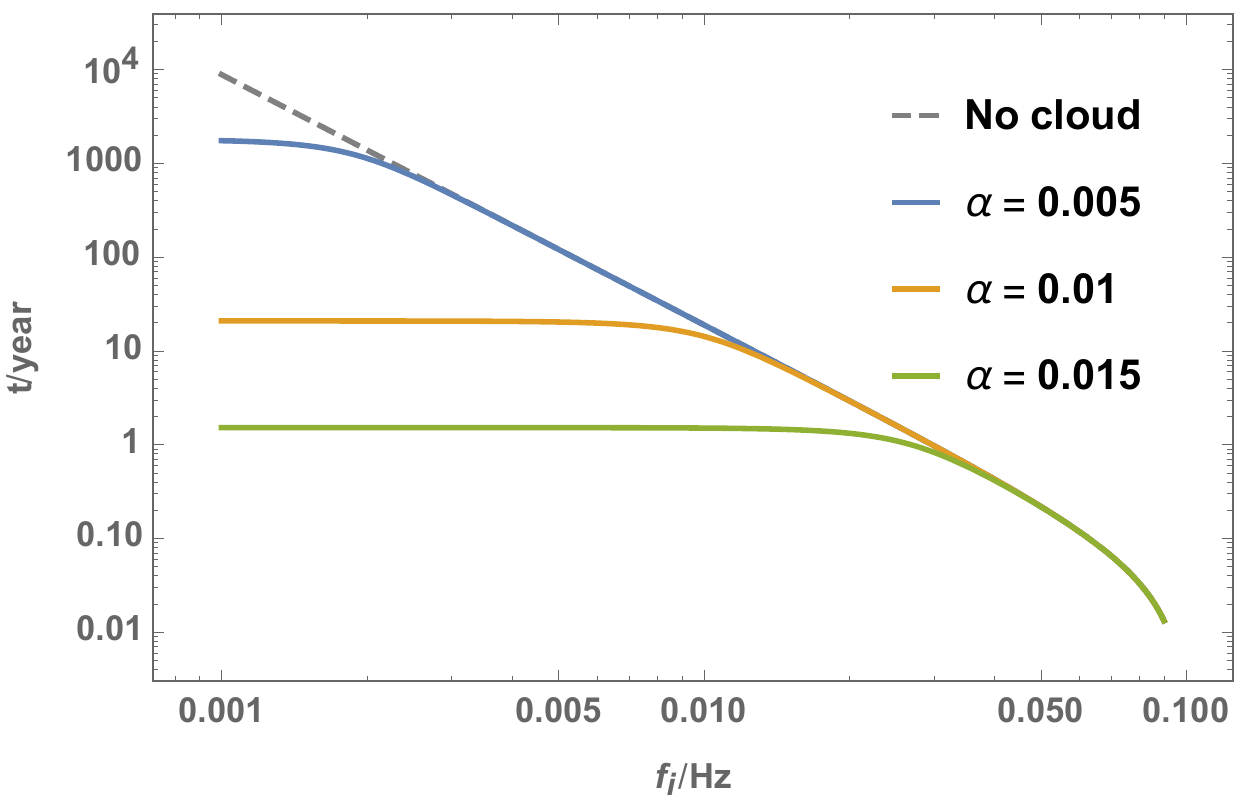} 
\caption{The timescales that the GW signal sweeps over the observation band of a LISA-like GW detector, i.e. $0.001 {\rm Hz} - 0.1 {\rm Hz}$. We assumed a binary of two $30 M_\odot$ BHs, and show the timescales with different initial frequency $f_i$, and different $\alpha$ (solid lines). We also show the case with no cloud (gray dashed lines) for comparison. In the upper panel, we assume no depletion of the cloud. In the lower panel, we assume the mass of the cloud is suppressed by a factor of $e^{-4}$ due to a previous depletion \cite{Baumann:2018vus,Berti:2019wnn}.} \label{fig:torb}
\end{figure}

 Let us consider a binary of two $30 M_\odot$ BHs, inspiraling in a circular orbit. The separation of the binary that corresponds to LISA's most sensitive observation band ($10^{-3}$Hz - 0.1 Hz) is from $2 \times 10^4 GM$ to $10^3 GM$. As a result, LISA may observe rapid orbit decay if the cloud radius is within such range, for example when $\alpha \sim 0.01$. At leading order, the chirp of GWs emitted by the orbit is
\ba
\frac{df}{dt} = -\frac{3}{\pi} \M^{-1} P,
\ea
where $\M \equiv \left(M M_*\right)^{3/5}/\left(M+M_*\right)^{1/5}$ is the chirp mass, and $P$ is the dissipation power of the orbital energy. Using $P = P_{\rm GW}$ as a benchmark, we can write
\ba
\frac{df}{dt} = \frac{96}{5} \pi^{8/3} \M^{5/3} f^{11/3} \left(1+\gamma\right),
\ea
where $\gamma$ is the fractional energy loss power caused by additional dissipation mechanisms other than the GW radiation, for example $\gamma_{\rm DF} = P_{\rm DF}/P_{\rm GW}$ for dynamical friction. The lifetime of the orbit can be estimated by
\ba\label{eq:torb}
\frac{t}{\M} = \int \frac{40}{3}\left(8\pi \M f\right)^{-11/3}\, \frac{8\pi \M \d  f}{1+\gamma}.  
\ea
In terms of observation, we are interested in the timescale that a GW signal sweeps over the observation band, given the initial frequency $f_i$ (or equivalently the initial separation of the binary). This timescale can be obtained by integrating Eq.~\eqref{eq:torb} from $f_i$ to $f_{\rm max}$ with $f_{\rm max}$ being the higher boundary of the observation band. In Fig.~\ref{fig:torb}, we show the timescales taking into account dynamical friction caused by the cloud. Note that the cloud could be depleted due to the level mixing happened earlier when the BHs were in a lower frequency orbit. In this case, the mass of the cloud could be suppressed at most by a factor of $e^{-4}$ for all $\alpha < 0.05$ \cite{Baumann:2018vus,Berti:2019wnn}. Nonetheless, the cloud is still able to accelerate the orbit decay significantly. 
In fact, we find that in most cases considered here the GW signals sweep through the observation band within one year, if $0.012 < \alpha < 0.04$ assuming no depletion of the cloud, or $0.016 < \alpha < 0.036$ assuming a depleted cloud.

\begin{figure}[tbp]
\centering 
\includegraphics[width=0.45\textwidth]{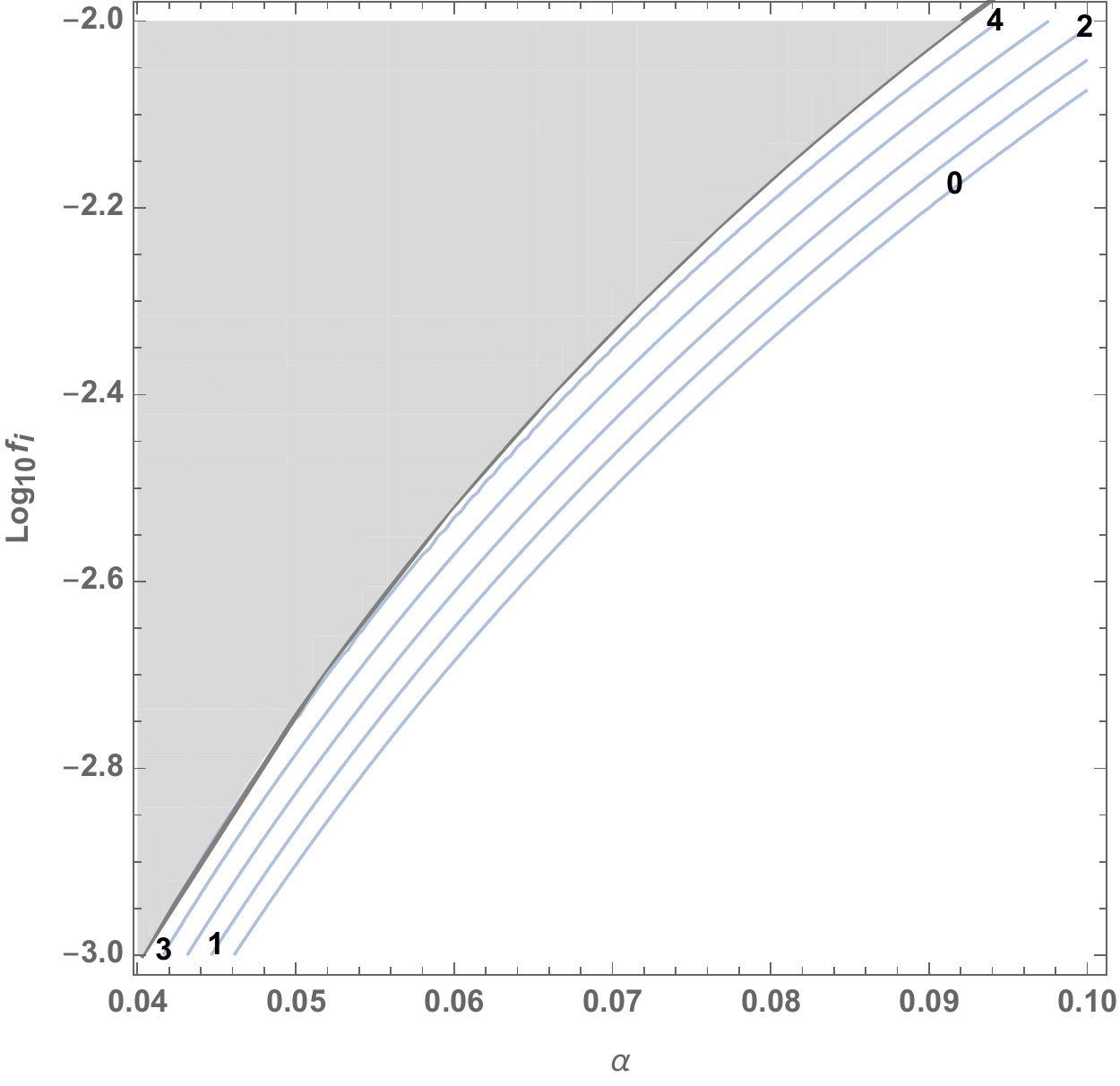} 
\includegraphics[width=0.45\textwidth]{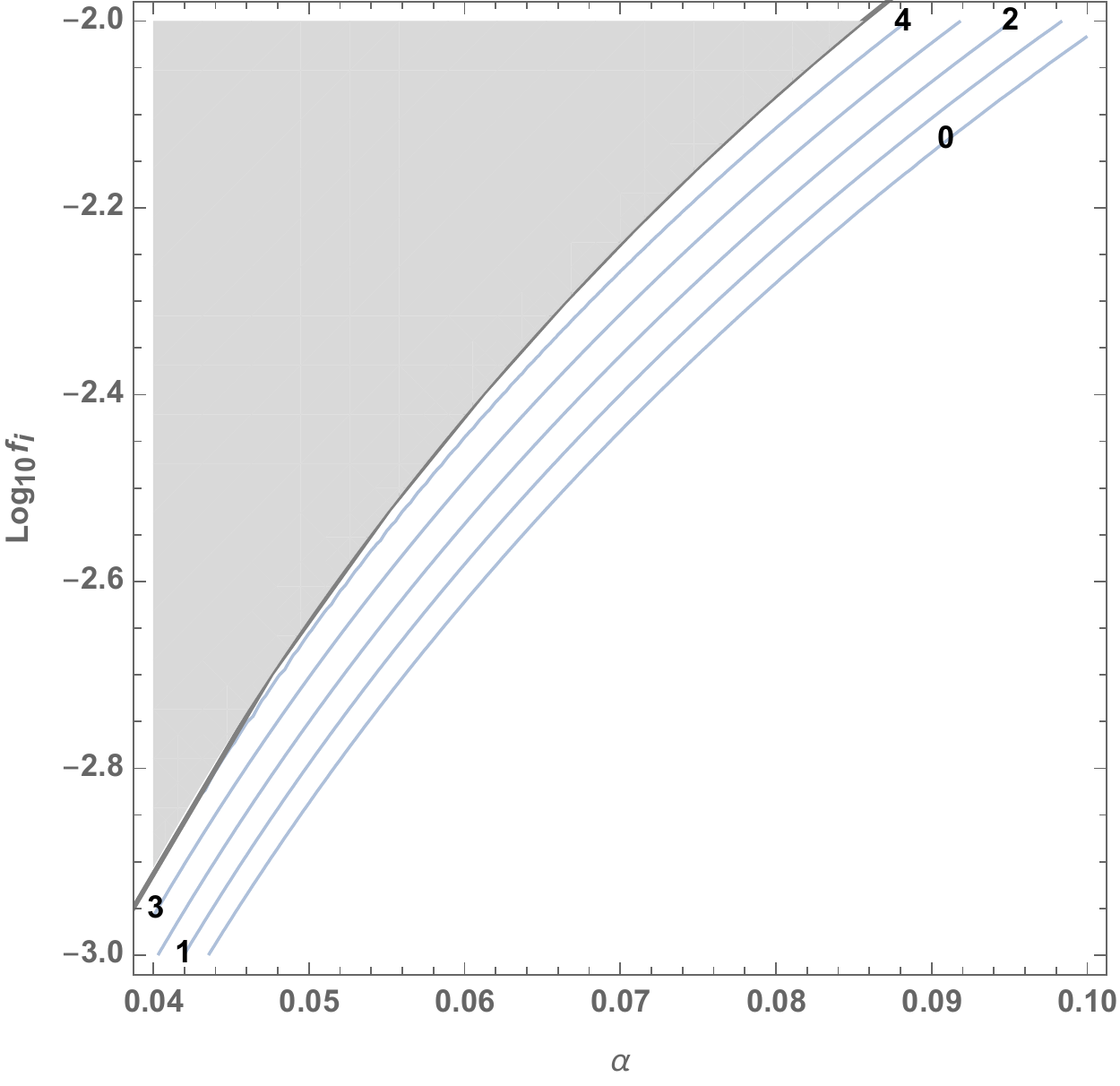} 
\caption{The phase difference $\Delta \Phi$ between the case with and without dynamical friction, assuming a binary of two $30 M_\odot$ BHs and one year integration time. In the gray region, the signal will sweep the observation band within one year taking into account the dynamical friction. The contours are labeled by $\log_{10} \Delta\Phi$. In the lower panel, we assume the mass of the cloud is suppressed by a factor of $e^{-4}$ due to a previous depletion \cite{Baumann:2018vus,Berti:2019wnn}.} \label{fig:phase}
\end{figure}

The detection of continuous signals is sensitive to the accumulated phase, which can be estimated by \cite{Cutler:1994ys}
\ba
\Phi = \int_{f_i}^{f_f} \frac{10}{3}\left(8\pi \M f\right)^{-8/3}\, \frac{8\pi \M \d  f}{1+\gamma}.
\ea
Here $f_i$ and $f_f$ are the initial and final frequency of the observation. In Fig.~\ref{fig:phase}, we calculated the phase difference between the case with and without dynamical friction. We assume one year integration time, and a binary of two $30 M_\odot$ BHs spiraling in a circular orbit. We can see that the phase difference caused by the cloud can be as large as $10^4$ when the dynamical friction is much stronger than the GW radiation reaction. Order $\mathcal{O}(1)$ phase shift is likely detectable by LISA parameter estimation procedures.

In principle, level mixing also affects the inspiral waveform. However, for stellar mass binaries, the effects of level mixing is either subdominated comparing to dynamical friction or happening outside the observation band.\footnote{Note that this is not true for EMRIs, in which case effects of level mixing is enhanced by $\eta^{-2}$ with $\eta$ being the small mass ratio.} Depending on the orbit frequency, level mixing can be dominated by Bohr resonance or hyperfine resonance \cite{Baumann:2018vus, Berti:2019wnn}. 
If the transient resonance timescale is longer than the observation time, one can use similar argument as in the dynamical friction case to estimate the phase modulation due to a continuous tidal force by mode mixing. On the other hand, if the transient resonance phase is much shorter than the observation time, the resonance essentially introduces a ``kick" in the orbital energy (see similar discussion in \cite{flanagan2012transient,yang2017general,bonga2019tidal}) that affects the accumulated phase later on.

Bohr resonance usually happens when the orbit size is comparable to that of the cloud, in which case the energy loss is usually dominated by dynamical friction. In particular, the resonance condition is that the orbital frequency $\Omega \simeq \Delta\omega/\Delta m \sim \mu \alpha^2 /\Delta m$, which corresponds to an orbital separation of $2^{1/3} \Delta m^{2/3} \alpha^{-2} GM$. In this case, $P_{\rm LM}/P_{\rm DF} \sim \alpha^{4\ell -1} \ll 1$. Comparing to Bohr resonance, hyperfine resonance could happen further away from the cloud, i.e. with a separation of $2^{1/3} \Delta m^{2/3} \alpha^{-4} GM$. In this case, we have $P_{\rm LM}/P_{\rm DF} \sim \alpha^{4\ell + 3} \I_{\rm DF} (x \sim \Delta 2^{1/3} m^{2/3} \alpha^{-2})$. As $\I_{\rm DF}$ decreases exponentially as $x$ increases, we may expect that for small $\alpha$, level mixing will eventually dominate dynamical friction. However, the corresponding frequency is usually outside LISA's observation band for such a small $\alpha$. For these reasons, we will not consider the effects of level mixing on the waveform.

\subsection{Binary formation rate}

Binary BHs may be born out of star binaries, or through dynamical multibody interactions in dense stellar environments, e.g., galactic nuclei where the number density can exceed $10^{10} {\rm pc}^{-3}$ \cite{OLeary:2008myb} in the inner region and globular clusters where the number density is about $10^5 {\rm pc}^{-3}$ \cite{Barack:2018yly}. In the LIGO O3 run new binary BH merger events are roughly detected in a weekly basis, where for third-generation detectors the event rate could be once per a few minutes \cite{abbott2017exploring}. With more detections it will be interesting to understand different possible formation channels. In the following, we will show that, in a certain range of velocity, the presence of superradiant clouds leads to a larger impact parameter for gravitational capture and hence a higher dynamical formation rate for binaries. It is particularly interesting if one or both BHs in the binary is(are) already product of previous mergers, because hierarchical mergers may be able to produce BHs heavier than the upper bound predicted by supernova pair instability \cite{belczynski2016effect,spera2017very,marchant2018pulsational,stevenson2019impact,woosley2017pulsational,gerosa2019escape}.

It would be convenient to write the loss of the orbit energy after one pericenter passage as
\ba\label{eq:Idef}
\Delta \e = \alpha^{3} \I,
\ea
where we approximately take $\I > 0.01$ for $1< x_p < 10$ due to dynamical friction (see Fig.~\ref{fig:Is}), and $\I > 0.001$ for $10<x_p<20$ due to level mixing (see Fig.~\ref{fig:de}). The angular momentum change caused by the cloud satisfies $\Delta J \ll J$ and will be neglected in the following discussion. Let us consider two BHs of comparable masses aiming at each other with an impact parameter $b$ and relative velocity $v_{\infty}$. In this case, the closest distance is
\ba
r_p &=& \left(\sqrt{\frac{1}{b^2}+\frac{M_{\rm tot}^2}{b^4v_{\infty}^4}}+\frac{M_{\rm tot}}{b^2v_{\infty}^2}\right)^{-1} \\
&\simeq& \frac{b^2v_{\infty}^2}{2 M_{\rm tot}}\left(1-\frac{b^2v_{\infty}^4}{4M_{\rm tot}^2}\right).
\ea
These two BHs will form a binary if they release enough energy during one passage, i.e. $\Delta \e > v_{\infty}^2/2$. For $\Delta \e \gg v_{\infty}^2/2$, the semi-major axis of new formed binary can be estimated by
\ba
a_0 \simeq \frac{M_{\rm tot}^2 \eta}{2 \Delta \e M_{\rm BH}},
\ea
and the eccentricity is
\ba
e_0 \simeq \sqrt{1- 2\frac{\Delta\e M_{\rm BH}b^2 v_{\infty}^2}{M_{\rm tot}^3\eta}} \simeq  \sqrt{1- 4\frac{\Delta\e M_{\rm BH}r_p}{M_{\rm tot}^2\eta}},
\ea
where we have used that $r_p \simeq b^2v_{\infty}^2/2 M_{\rm tot}$. As $\Delta \e \ll 1$, we have $e_0 \simeq 1$.
In addition, a successful merger requires that the newly formed binary should not be disrupted by further gravitational encounter with a third body within the merger timescale. The lifetime of a highly eccentric ($e \simeq 1$) orbit (due to GW emission) is given by \cite{Peters:1963ux}
\ba
t_0 \simeq \frac{3}{85} \frac{a_0^4}{M_{\rm tot}^3\eta}(1-e_0^2)^{7/2},
\ea
while the typical timescale for an encounter to disrupt the binary is \cite{OLeary:2008myb}
\ba
t_e \simeq \frac{1}{12\pi a_0^2 n_{\rm BH} v} \simeq \frac{v^3}{12\pi M_{\rm tot}^2 n_{\rm BH}},
\ea
where $n_{\rm BH}$ is the number density of the BHs, and $v$ is the typical velocity of the binary. If the energy dissipation is dominated by dynamical friction or level mixing, we have
\ba
\frac{a_0}{M_{\rm tot}} \simeq \frac{M_{\rm tot}\eta}{2 M} \alpha^{-3}\I^{-1},
\ea
and
\ba
1-e_0^2 \simeq \frac{4M}{M_{\rm tot}\eta} \alpha \I x_p.
\ea
Therefore, we have
\ba
\frac{t_0}{t_e} = 2\times10^{-3} \I^{-1/2} \left(\frac{x_p}{10}\right)^{7/2}\left(\frac{v}{100{\rm kms^{-1}}}\right)^{-3} \nonumber \\
\times \left(\frac{\alpha}{0.1}\right)^{-17/2} \left(\frac{M}{30M_\odot}\right)^{3}  \left(\frac{n_{\rm BH}}{10^8 {\rm pc}^{-3}}\right),
\ea
which means $t_0 \ll t_e$ if the BH number density is not extreme large. For example, we can find that for $v = 100 {\rm km} {\rm s}^{-1}$, in order for $t_0 \sim t_e$, the number density of BHs must be $n_{\rm BH} \ge 2\times 10^{10}\, {\rm pc}^{-3}$ for $\alpha = 0.1$ or $n_{\rm BH} \ge 8\times 10^{8}\, {\rm pc}^{-3}$ for $\alpha = 0.05$. As long as the BH number density is below the corresponding value, the rate of merger is roughly the rate of binary formation.

For simplicity, we simplify the mass distribution of BHs by assuming all BHs have the same mass, in which case the binary formation rate can be estimated as
\ba\label{eq:GBF}
\Gamma_{\rm BF} = \int n_{BH}^2 4\pi^2 b_{\rm max}^2 v_{\infty} r^2 dr,
\ea
where $b_{\rm max}$ is the maximum impact parameter for two BHs with relative velocity $v_{\infty}$ to be bound. For instance, in the galactic nuclei, we have $n_{\rm BH} \propto r^{\beta}$, where $r$ is the distance to the galaxy centre, and $v_{\infty} \sim \sqrt{GM_{\rm SMBH}/r}$ with $M_{\rm SMBH}$ being the mass of the supermassive BH in the centre of the galaxy. For globular clusters, $v_{\infty}$ is given by the typical velocity dispersion. If we are only interested in binaries that lead to successful mergers, the integral~\eqref{eq:GBF} should have a lower cutoff at $r$ such that $t_0 \simeq t_e$. This cutoff may be relevant for binary formation in galactic nuclei, where the BH number density in the inner region could be too large for new formed binaries to survive. On the other hand, it may not be a problem in globular clusters where the BH number density is usually much lower.

Without superradiant cloud, BHs can form binaries by GW dissipation, which defines a maximum impact parameter \cite{OLeary:2008myb}
\ba\label{eq:bGW}
b_{\rm GW}= \left(\frac{340\pi}{3}\right)^{1/7} \eta^{1/7} M_{\rm tot} v_{\infty}^{-9/7}.
\ea
Similarly, we can also find the maximum impact parameter if the dissipation is caused by the cloud
\ba\label{eq:bcloud}
b_{\rm cloud} \simeq \sqrt{ \frac{2  \alpha^{-2}x_p M_{\rm tot} }{v_{\infty}^2 M} }.
\ea
As the dissipation caused by cloud becomes significant only when the periapsis is comparable to the size of the cloud, we can choose $x_p \sim 10$ for $v_{\infty} < 500 {\rm km}{\rm s}^{-1}$ with $\alpha$ ranged from $0.01$ to $0.1$. We see that the maximum impact parameter becomes larger in the presence of superradiant clouds. For example, we have $b_{\rm max}^{\rm DF} = 10\, b_{\rm max}^{\rm GW}$ for $\alpha = 0.01$ and $v_{\infty} = 100 {\rm km}{\rm s}^{-1}$.

Beside the formation dynamics, binary formation rate also depends on the environments, such as the BH distribution in the galactic nuclei. Therefore, we shall compare the formation rate in the presence of cloud with that associated with GW radiation, by computing $\gamma \equiv \Gamma_{\rm BF}^{\rm cloud}/\Gamma_{\rm BF}^{\rm GW}$. In galactic nuclei, integrating Eq.~\eqref{eq:GBF} with Eq.~\eqref{eq:bGW} and Eq.~\eqref{eq:bcloud}, gives
\ba\label{eq:rGN}
\gamma_{\rm GN} \simeq 56\, C_\beta\, \left(\frac{x_p}{10}\right) \left(\frac{\alpha}{0.01}\right)^{-2}  \left(\frac{M_{\rm SMBH}}{10^6 M_\odot}\right)^{2/7},
\ea
where
\ba
C_\beta =\frac{\int_{x^{\rm cloud}_{\rm min}}^{x_{\rm max}}dx\, x^{2\beta + \frac{5}{2}}}{\int_{x^{\rm GW}_{\rm min}}^{x_{\rm max}} dx\, x^{2\beta+\frac{39}{14}}}
\ea
is an order one coefficient depends on the slope of the BH number density $\beta$. For instance, we have $C_\beta \simeq 0.7$ for $\beta = -2$ \cite{Hopman:2006qr}, and $C_\beta \simeq 1.6$ for $\beta = -7/4$ \cite{Bahcall:1976aa}. For binary formation in globular clusters, both number density $n_{\rm BH}$ and $v_{\infty}$ are smaller than that in galactic nuclei, therefore we have 
\ba\label{eq:rGC}
\gamma_{\rm GC} &=& \frac{b_{\rm cloud}^2}{b_{\rm GW}^2} \nonumber \\
&\simeq&213 \left(\frac{x_p}{10}\right) \left(\frac{\alpha}{0.01}\right)^{-2} \left(\frac{v_{\infty}}{60 {\rm kms}^{-1}}\right)^{4/7}.
\ea
Eq.~\eqref{eq:rGN} and Eq.~\eqref{eq:rGC} indicate that, in the presence of superradiant cloud, the binary formation rate will be significantly enhanced if $\alpha <  0.1$.

In the above estimation, we have assumed that all BHs have the same mass. In the realistic case, we may expect BH mass ranges from $\sim 5 M_{\odot}$ to $\sim 50 M_{\odot}$. Given the mass of the scalar field, BHs of different mass should be surrounded with cloud of different $\alpha$. According to Eq.~\eqref{eq:rGN} and Eq.~\eqref{eq:rGC}, we may expect that small BHs are more like to form binaries comparing to large BHs. The estimation can also be improved by considering a more accurate velocity distribution instead of $v_{\infty} \sim \sqrt{GM_{\rm SMBH}/r}$. A detailed modeling is beyond the scope of this paper, and will be left for future works.

\section{Discussion}
\label{sec:Dis}

Superradiant cloud may develop around a rotating (astrophysical) BH, if the Compton wavelength of the boson field is comparable to the size of the BH. In this paper, we investigate the effects of the superradiant cloud of a scalar field on the orbits of nearby compact objects, focusing on dynamical friction and backreaction of cloud level mixing. We compute the dynamical friction caused by the cloud, see Eq.~\eqref{eq:deDF}. Depending on the mass of the scalar field, the dynamical friction dissipation can be larger than the dissipation caused by GW radiation. Therefore the dynamical friction dissipation is able to accelerate orbit decay, leaving fingerprints in the GW waveform. For EMRIs around super-massive BHs, it is not very likely to detect the effects of dynamical friction in the EMRI waveform, as a LISA-like detector usually see inspirals with separations much less than the size of the cloud and the effects of dynamical friction is not maximized in the observation band. The effects are much more significant on stellar mass BH inspirals. For example, with $0.016 < \alpha < 0.036$, GW signals from a binary of two $30 M_\odot$ BHs can sweep over the whole LISA band within one year, which usually takes $10^4$ year in the absence of the cloud. For $\alpha > 0.036$, the cloud could also introduce a large phase difference that is up to $10^4$ assuming one year observation. However, for EMRIs around intermedia mass BHs, there could be interesting GW signals that can be observed by LISA \cite{Zhang:2018kib}. In addition to dynamical friction, when a compact object approaches the cloud, it could dynamically excite the eigenmodes of the cloud. In most cases, the object will lose energy to the cloud due to the tidal interaction. We calculate the energy loss during one passage by considering a parabolic orbit, see Fig.~\ref{fig:de}. 

In both cases, we expect that the dissipation caused by the cloud will facilitate the cloudy BH capturing other objects, such as BHs and neutron stars. We estimate the effects of the superradiant cloud on the EMRI rate, and find the EMRI rate does not change very much in the presence of the cloud. It is because the EMRI rate is mainly determined by repopulation mechanism of the loss cone. We also investigate the formation rate of stellar mass BH binaries. Assuming a single mass distribution of BHs, we find the formation rate is enhanced in the presence of superradiant cloud, if $\alpha < 0.1$. All of these effects provide promising ways of searching for light bosons.

A large binary formation rate could lead to interesting observational signatures. One possible consequence is hierarchical mergers, i.e. mergers with one BH being the remnant of a previous merger. If there is a scalar field of mass $10^{-11} {\rm eV}$, we would have cloud developing around stellar mass BHs. The growth time of superradiant cloud with $\alpha \ll 1$ can be estimated as \cite{Detweiler:1980uk}
\ba\label{eq:gtime}
\tau \simeq 24 (a/M)^{-1} \alpha^{-\left(4\ell+5\right)} \left(GM/c^3\right),
\ea
where $a$ is the spin of the BH. For $30 M_\odot$ BHs, the growth time for $\ell = 1$ mode is about $10^8$ years if $\alpha = 0.01$ or equivalently if there is a scalar field of mass $\mu = 4.5 \times 10^{-11} {\rm eV}$. Obviously in order to be astrophysically relevant, according to Eq.~\eqref{eq:gtime} $\alpha$ cannot be much smaller than $0.01$, or the growth timescale will be longer than the Hubble time. With $\alpha \ge 0.01$ the cloud may develop around the remnant BHs of previous mergers and enhance the formation rate of binaries with higher generation BHs. 

On the other hand, such rate enhancement mechanism may play a role in the growth of supermassive BHs, e.g. QSO SDSS 1148+5251 found at $z=6.43$ and with mass $\sim 10^9 M_\odot$ \cite{fan2003survey}. It remains an open question that how do the supermassive BHs grow out of their much lighter seeds. Gas accretion may not be able to support sufficiently fast mass growth \cite{natarajan2019disentangling}, so that BH merger is expected to play a role. The enhanced capture rate with scalar cloud may allow fast growth of the host BH across one order of magnitude.

\acknowledgements

We thank Sam Dolan for helping with the accretion cross section of a BH in scalar cloud. J.Z. is supported by European Union's Horizon 2020 Research Council grant 724659 MassiveCosmo ERC-2016-COG. H.Y.~acknowledges support from the Natural Sciences and Engineering Research Council of Canada, and in part by the Perimeter Institute for Theoretical Physics. Research at Perimeter Institute is supported by the Government of Canada through the Department of Innovation, Science and Economic Development Canada, and by the Province of Ontario through the Ministry of Research and Innovation. 

\appendix

\section{Supperradiance cloud in non-relativistic limit}
\label{app:SC}

In this appendix, we summarize the non-relativistic description of a superradiant cloud. The Lagrangian of a free complex scalar field $\Psi$ is
\ba\label{app:L}
{\cal L} = - g^{ab}\partial_a \Psi^* \partial_b \Psi - \mu^2 \Psi^* \Psi,
\ea
where $g_{ab}$ is the metric of Kerr BH, assuming the backreaction of the field's stress-energy on the metric is negligible. One can make an ansatz
\ba
\Psi (t, {\bf r}) = \frac{1}{\sqrt{2\mu}} \psi(t, {\bf r}) e^{-i\mu t},
\ea
where $\psi$ is a complex scalar field which varies on a timescale much longer than $\mu^{-1}$. Given Lagrangian \eqref{app:L}, the action of $\psi$ is 
\ba
&S =-\frac{1}{2 \mu}\int \d^4 x \sqrt{-g} \Big[ \partial_a \psi^* \partial^a \psi   + \mu^2 ( g^{00} + 1)  \psi^* \psi \nonumber \\
& + i \mu g^{0 a}  \left( \psi^* \partial_a \psi -  \psi \partial_a \psi^* \right) \Big].
\ea 
Keeping only terms up to first order in $r^{-1}$ and $\alpha^2$, we can obtain the effective action,
\ba
S_{2} = \int  dt\, d^3{\bf r}\left[ \psi^* \partial_t \psi - \frac{1}{2 \mu }  {\bf \nabla} \psi^*  {\bf \nabla} \psi + \frac{\alpha}{r}  \psi^* \psi  \right], 
\ea
which leads to the Schr\"odinger equation
\ba
i \frac{\partial }{\partial t} \psi(t, {\bf r}) = \left[ - \frac{1}{2\mu} \nabla^2 - \frac{\alpha}{r} \right] \psi(t, {\bf r}) \, .
\ea
The stationary eigenmode of cloud is denoted by $\ket{n\ell m}$ with the wave function
\ba
\psi_{n\ell m} = e^{-i\left(\omega - \mu \right)t}\, R_{n\ell} \left(x\right) \,Y_{\ell m} (\theta,\, \phi),
\ea
where $x = \alpha^2 r/GM$, $Y_{\ell m}$ is the spherical harmonic function, and
\ba
R_{n\ell}(x) &=& \left[\left(\frac{2}{n}\right)^3 \frac{\left(n-\ell-1\right)!}{2n(n+1)!}\right]^{1/2} e^{-\frac{x}{n}}\left(\frac{2x}{n}\right)^\ell \nonumber \\
&&\times L_{n-\ell+1}^{2\ell+1} \left[\frac{2x}{n}\right]
\ea
with $ L_{n-\ell+1}^{2\ell+1}[x]$ being the generalized Laguerre polynomial of degree $n-\ell-1$. 

\section{Tidal perturbation}
\label{app:tide}

In this appendix, we show the explicit expression of $V_*$. In the frame centred at the BH, the Hamiltonian of the system is
\ba\label{eq:Htot}
H_{\rm tot} &=& \left[\frac{{\bf p}^2}{2\mu} - \frac{\mu M}{r} \right]  \nonumber \\
&+& \left[ \frac{{\bf p}_*^2}{2M_*} - \frac{M_*M}{R_*} - \frac{M_*\mu}{\left| {\bf R}_* - {\bf r} \right|} + \frac{M_*\mu}{R_*^3} {\bf r}\cdot {\bf R}_*\right], \nonumber \\
 \ea
where ${\bf p} = \mu \dot{{\bf r}}$, ${\bf p}_* = M_* \dot{{\bf R}}_*$ with ${\bf r}$ and ${\bf R_*}$ being the positions of the cloud and the star relative to the BH. For $r < R_*$, we have
\ba\label{eq:1overR}
 \frac{1}{\left| {\bf R}_* - {\bf r} \right|}  = \frac{1}{R_*} + \frac{r \cos \Delta\theta}{R_*^2}  +  \sum_{\ell_* \ge 2}\, \sum_{\left|m_*\right| \le \ell_*} \frac{4\pi}{2\ell_*+1}\nonumber \\ 
\times \frac{r^{\ell_*}}{R_*^{\ell_*+1}} Y^*_{\ell_* m_*}\left(\theta_*,\phi_*\right) Y_{\ell_*  m_*}\left(\theta,\phi\right),
\ea
where $\Delta \theta$ is the angle between ${\bf r}$ and ${\bf R}_*$. The first term on the r.h.s. of Eq.~\eqref{eq:1overR} does not change the eigenstate of the cloud, and the second term cancels with the last term in Eq.~\eqref{eq:Htot}. For $R_* < r$, we have
\ba
 \frac{1}{\left| {\bf R}_* - {\bf r} \right|}  = \frac{1}{r} + \frac{{\bf r}\cdot {\bf R}_*}{r^3} + \sum_{\ell_* \ge 2} \,\sum_{\left|m_*\right| \le \ell_*} \frac{4\pi}{2\ell_*+1} \nonumber \\
 \times \frac{R_*^{\ell_*}}{r^{\ell_*+1}} Y^*_{\ell_* m_*}\left(\theta_*,\phi_*\right) Y_{\ell_*  m_*}\left(\theta,\phi\right),
\ea
in which case the monopole and dipole terms contribute. 

The inner product can be written as
\ba
\left<\psi_i \left|V_*\right|\psi_j\right> &=& -M_* \mu \sum_{\ell_*,\, m_*} \frac{4\pi}{2\ell + 1} \,\I_\Omega\, \I_r(r_*) Y_{\ell_*m_*}(\theta_*,0) \nonumber \\
&&\times \exp \left[ i(\omega_i-\omega_j)t\mp im_* \phi_*\right],
\ea
where the upper sign in front of $m_*\phi_*$ corresponds to co-rotating orbits, while the lower sign corresponds to counter-rotating orbits. For $\ell_* \ge 2$, we have
\ba
\I_{r}=\int_0^{\infty} dr \, r^2 \frac{r_{<}^{\ell_*}}{r_{>}^{\ell_*+1}} R_{n_i \ell_i}\left(x\right) R_{n_j \ell_j}\left(x\right),
\ea
where $x\equiv\alpha^{-2}r/GM$, $r_{<}$ is the smaller of $r$ and $r_*$ and $r_{>}$ is the lager of $r$ and $r_*$. For $\ell_* = 1$ and $0$, we have
\ba
\I_{r} = \int_{r_*}^{\infty} dx \, r_* \left(1- \frac{r^3}{r_*^3}\right) R_{n_i \ell_i}\left(x\right) R_{n_j \ell_j}\left(x\right)
\ea
and 
\ba
\I_{r} = \int_{r_*}^{\infty} dr \, r R_{n_i \ell_i}\left(x\right) R_{n_j \ell_j}\left(x\right) 
\ea 
respectively.

\section{Effects of mode decay and floating orbits}
\label{app:MD}

In this appendix, we will discuss the effects of mode decay. In particular, we will show that for EMRIs with quasi-circular orbits, dynamical friction dominates over the backreation of level mixing during most of the time, while the floating orbits proposed in \cite{Zhang:2018kib} are still possible at resonance frequency.
\begin{figure}[tbp]
\centering 
\includegraphics[width=0.45\textwidth]{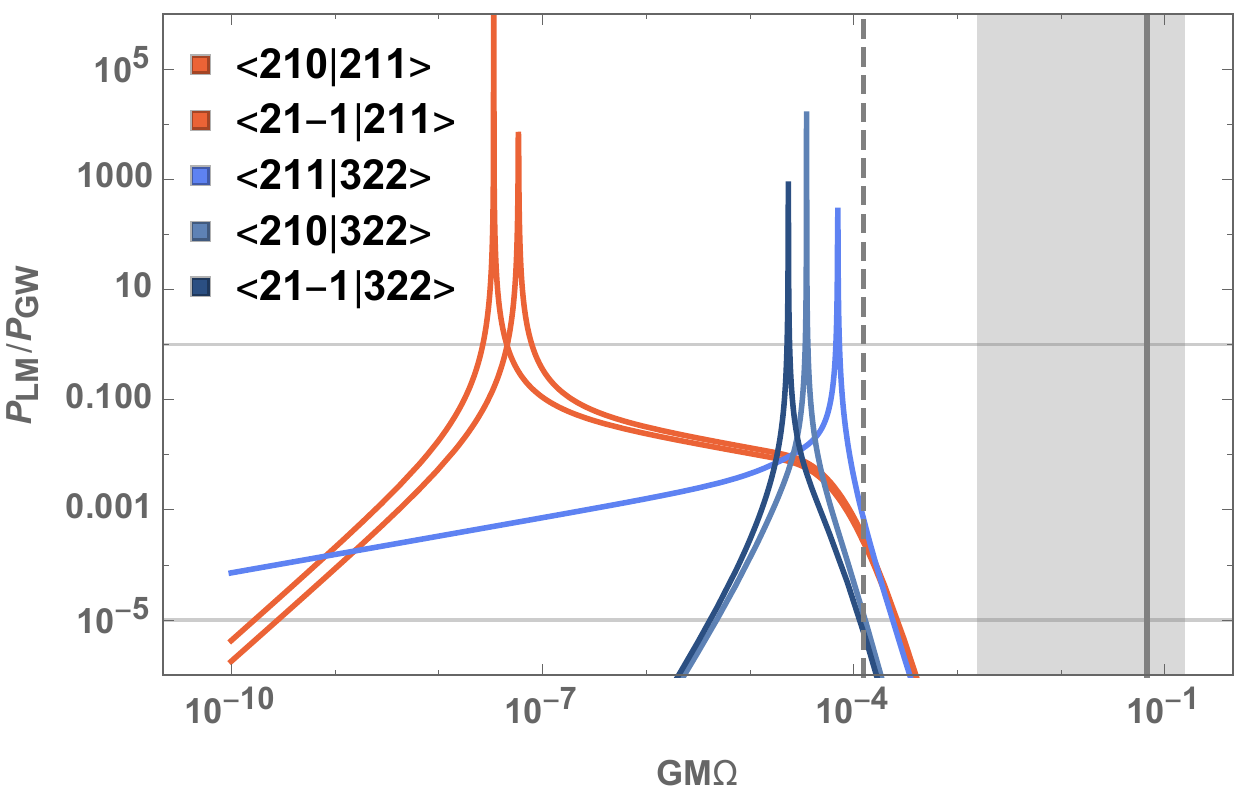} 
\includegraphics[width=0.45\textwidth]{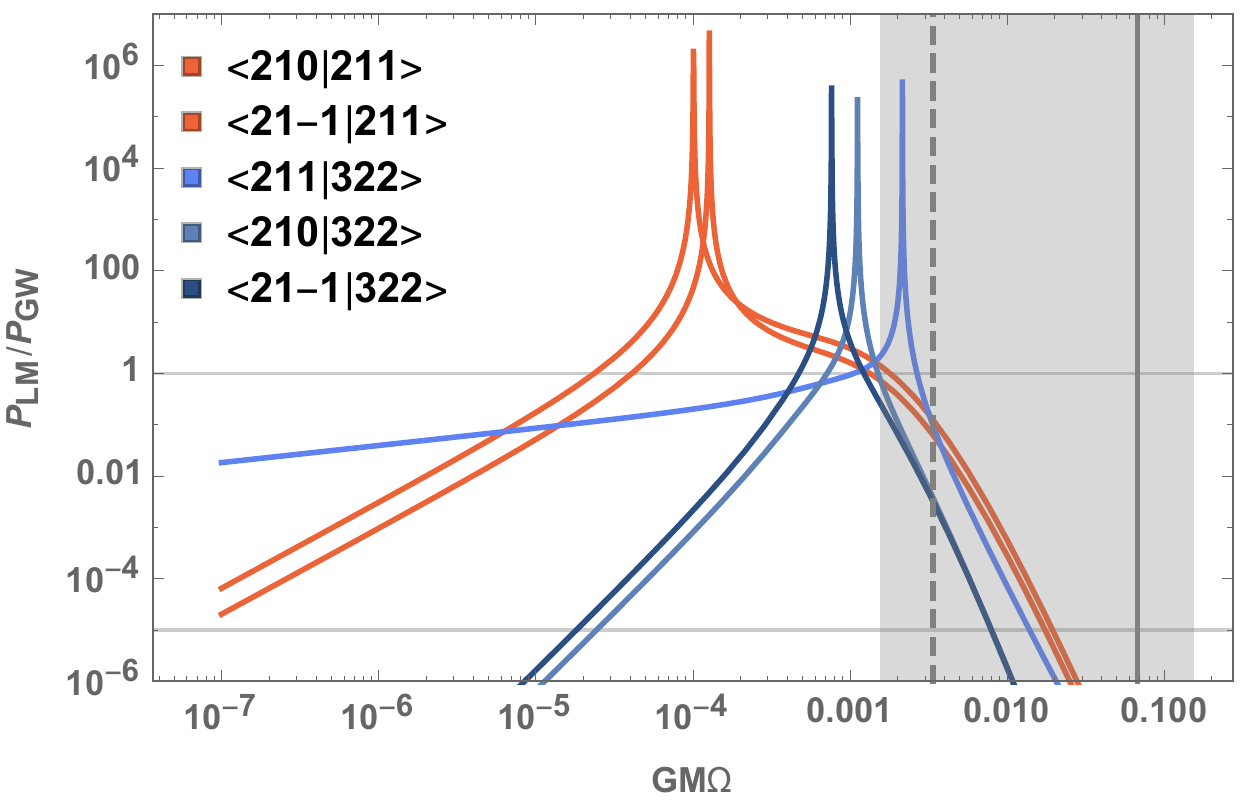} 
\caption{The ratio between angular momentum changing rate caused by level mixing and that caused by GW radiation. We consider a circular orbit with the inclination $i=\pi/4$ (as some coupling strength is proportional to $\cos i$ or $\sin i$), and show the ratio for the dominant level mixing with $\ket{211}$ and $\ket{322}$. $\alpha = 0.1$ for the upper plot and $\alpha = 0.3$ for the lower plot. The gray dashed line shows the orbital frequency corresponding to a radius $4\alpha^{-2}$, which locates the density peak of the cloud. The shaded region shows the LISA frequency band ($0.1$Hz - $0.001$Hz), assuming a $10^5 M_{\odot}$ BH. The gray vertical line denotes the innermost stable circular orbit.} \label{fig:LM}
\end{figure}

The tidal field of the star can mix the modes of the cloud, which leads to angular momentum transfer between the star and the BH-cloud system. For a quasi-circular orbit, the angular momentum transfer has been calculated in \cite{Zhang:2018kib},
\ba\label{eq:transferrate}
\left< \frac{dL_{*}}{dt}\right> = \Delta m \Gamma\frac{Q^2}{2\omega_{\rm R}^2} \frac{M_c}{\mu},
\ea
where $\Gamma$ is the decay rate of the mode that the saturated mode coupling to, $Q$ is the coupling strength, $\omega_R$ is the Rabi frequency defined as
\ba
\omega_{\rm R} = \sqrt{Q^2 + \left(\Delta\omega - \Delta m \Omega\right)^2/4},
\ea
where $\Delta \omega$ and $\Delta m$ are the energy level and quantum number difference between the two modes. On the other hand, the angular momentum carried away by GW radiation is
\ba
\left. \left< \frac{dL}{dt}\right> \right |_{\rm GW} = -\frac{32}{5}\eta^2 M \left(GM\Omega\right)^{7/3}. 
\ea
Fig.~\ref{fig:LM} shows the relative magnitude of angular momentum changing caused by level mixing and GW radiation. Floating orbits occur when the angular momentum loss by GW radiation is compensated by that gained through level mixing. When there is dynamical friction, the occurrence of floating orbits also requires energy gained from level mixing can balance the dissipation caused by dynamic friction. The maximum power of energy gaining due to level mixing, comparing GW radiation is
\ba
\left. \frac{P_{\rm LM}}{P_{\rm GW}} \right|_{\rm max}\sim \frac{5}{64} \Delta m^{10/3} \eta^{-2} \alpha^{p_\ell}, 
\ea
where $p_\ell = 4\ell-1$ for Bohr resonance and $p_\ell = 4\ell-8$ for hyperfine resonance. On the other hand, the ratio between the effects of GW is given by Eq.~\eqref{eq:DFGWratio}, comparing to which we have
\ba
\frac{\left. P_{\rm LM} \right|_{\rm max}}{P_{\rm DF}} > \frac{\left. P_{\rm LM} \right|_{\rm max}}{\left.P_{\rm DF} \right|_{\rm max}}\sim 10^9 \Delta m^{\frac{10}{3}}\, \alpha^{p_\ell+4}  \left(\frac{\eta}{10^{-5}}\right)^{-2}. \nonumber \\
\ea

For orbits around a super-massive BH, we usually have $\eta \ll 10^{-5}$, in which case the energy gain due to the coupling to a decaying mode can still compensate the energy loss caused dynamical friction. For orbits around an intermedia mass BH, the energy gain due to the coupling to a decaying mode may no longer support the orbital decay in the presence of dynamical friction. However, a floating orbit still exists if one considers the time dependence of the orbital frequency as discussed in \cite{Baumann:2019ztm}. In the presence of the dynamical friction, the change of the orbital frequency near the resonance frequency may be dominated by dynamical friction. Nevertheless, the Landau-Zener parameter defined in \cite{Baumann:2019ztm} is still much greater than $1$. Therefore, one can still expect long-lasting monochromatic GW signals from floating orbits around intermedia mass BHs as predicted in \cite{Zhang:2018kib}.

\bibliography{master}

\begin{thebibliography}{82}
\expandafter\ifx\csname natexlab\endcsname\relax\def\natexlab#1{#1}\fi
\expandafter\ifx\csname bibnamefont\endcsname\relax
  \def\bibnamefont#1{#1}\fi
\expandafter\ifx\csname bibfnamefont\endcsname\relax
  \def\bibfnamefont#1{#1}\fi
\expandafter\ifx\csname citenamefont\endcsname\relax
  \def\citenamefont#1{#1}\fi
\expandafter\ifx\csname url\endcsname\relax
  \def\url#1{\texttt{#1}}\fi
\expandafter\ifx\csname urlprefix\endcsname\relax\def\urlprefix{URL }\fi
\providecommand{\bibinfo}[2]{#2}
\providecommand{\eprint}[2][]{\url{#2}}

\bibitem[{\citenamefont{Peccei and Quinn}(1977)}]{PhysRevLett.38.1440}
\bibinfo{author}{\bibfnamefont{R.~D.} \bibnamefont{Peccei}} \bibnamefont{and}
  \bibinfo{author}{\bibfnamefont{H.~R.} \bibnamefont{Quinn}},
  \bibinfo{journal}{Phys. Rev. Lett.} \textbf{\bibinfo{volume}{38}},
  \bibinfo{pages}{1440} (\bibinfo{year}{1977}).

\bibitem[{\citenamefont{Weinberg}(1978)}]{PhysRevLett.40.223}
\bibinfo{author}{\bibfnamefont{S.}~\bibnamefont{Weinberg}},
  \bibinfo{journal}{Phys. Rev. Lett.} \textbf{\bibinfo{volume}{40}},
  \bibinfo{pages}{223} (\bibinfo{year}{1978}).

\bibitem[{\citenamefont{Turner}(1983)}]{PhysRevD.28.1243}
\bibinfo{author}{\bibfnamefont{M.~S.} \bibnamefont{Turner}},
  \bibinfo{journal}{Phys. Rev. D} \textbf{\bibinfo{volume}{28}},
  \bibinfo{pages}{1243} (\bibinfo{year}{1983}).

\bibitem[{\citenamefont{Press et~al.}(1990)\citenamefont{Press, Ryden, and
  Spergel}}]{PhysRevLett.64.1084}
\bibinfo{author}{\bibfnamefont{W.~H.} \bibnamefont{Press}},
  \bibinfo{author}{\bibfnamefont{B.~S.} \bibnamefont{Ryden}}, \bibnamefont{and}
  \bibinfo{author}{\bibfnamefont{D.~N.} \bibnamefont{Spergel}},
  \bibinfo{journal}{Phys. Rev. Lett.} \textbf{\bibinfo{volume}{64}},
  \bibinfo{pages}{1084} (\bibinfo{year}{1990}).

\bibitem[{\citenamefont{Hu et~al.}(2000)\citenamefont{Hu, Barkana, and
  Gruzinov}}]{Hu:2000ke}
\bibinfo{author}{\bibfnamefont{W.}~\bibnamefont{Hu}},
  \bibinfo{author}{\bibfnamefont{R.}~\bibnamefont{Barkana}}, \bibnamefont{and}
  \bibinfo{author}{\bibfnamefont{A.}~\bibnamefont{Gruzinov}},
  \bibinfo{journal}{Phys. Rev. Lett.} \textbf{\bibinfo{volume}{85}},
  \bibinfo{pages}{1158} (\bibinfo{year}{2000}), \eprint{astro-ph/0003365}.

\bibitem[{\citenamefont{Peebles}(2000)}]{Peebles:2000yy}
\bibinfo{author}{\bibfnamefont{P.~J.~E.} \bibnamefont{Peebles}},
  \bibinfo{journal}{Astrophys. J.} \textbf{\bibinfo{volume}{534}},
  \bibinfo{pages}{L127} (\bibinfo{year}{2000}), \eprint{astro-ph/0002495}.

\bibitem[{\citenamefont{Amendola and Barbieri}(2006)}]{Amendola:2005ad}
\bibinfo{author}{\bibfnamefont{L.}~\bibnamefont{Amendola}} \bibnamefont{and}
  \bibinfo{author}{\bibfnamefont{R.}~\bibnamefont{Barbieri}},
  \bibinfo{journal}{Phys. Lett.} \textbf{\bibinfo{volume}{B642}},
  \bibinfo{pages}{192} (\bibinfo{year}{2006}), \eprint{hep-ph/0509257}.

\bibitem[{\citenamefont{Schive et~al.}(2014)\citenamefont{Schive, Chiueh, and
  Broadhurst}}]{Schive:2014dra}
\bibinfo{author}{\bibfnamefont{H.-Y.} \bibnamefont{Schive}},
  \bibinfo{author}{\bibfnamefont{T.}~\bibnamefont{Chiueh}}, \bibnamefont{and}
  \bibinfo{author}{\bibfnamefont{T.}~\bibnamefont{Broadhurst}},
  \bibinfo{journal}{Nature Phys.} \textbf{\bibinfo{volume}{10}},
  \bibinfo{pages}{496} (\bibinfo{year}{2014}), \eprint{1406.6586}.

\bibitem[{\citenamefont{Hui et~al.}(2017)\citenamefont{Hui, Ostriker, Tremaine,
  and Witten}}]{Hui:2016ltb}
\bibinfo{author}{\bibfnamefont{L.}~\bibnamefont{Hui}},
  \bibinfo{author}{\bibfnamefont{J.~P.} \bibnamefont{Ostriker}},
  \bibinfo{author}{\bibfnamefont{S.}~\bibnamefont{Tremaine}}, \bibnamefont{and}
  \bibinfo{author}{\bibfnamefont{E.}~\bibnamefont{Witten}},
  \bibinfo{journal}{Phys. Rev.} \textbf{\bibinfo{volume}{D95}},
  \bibinfo{pages}{043541} (\bibinfo{year}{2017}), \eprint{1610.08297}.

\bibitem[{\citenamefont{Marsh}(2016)}]{Marsh:2015xka}
\bibinfo{author}{\bibfnamefont{D.~J.~E.} \bibnamefont{Marsh}},
  \bibinfo{journal}{Phys. Rept.} \textbf{\bibinfo{volume}{643}},
  \bibinfo{pages}{1} (\bibinfo{year}{2016}), \eprint{1510.07633}.

\bibitem[{\citenamefont{Arvanitaki et~al.}(2010)\citenamefont{Arvanitaki,
  Dimopoulos, Dubovsky, Kaloper, and March-Russell}}]{arvanitaki2010string}
\bibinfo{author}{\bibfnamefont{A.}~\bibnamefont{Arvanitaki}},
  \bibinfo{author}{\bibfnamefont{S.}~\bibnamefont{Dimopoulos}},
  \bibinfo{author}{\bibfnamefont{S.}~\bibnamefont{Dubovsky}},
  \bibinfo{author}{\bibfnamefont{N.}~\bibnamefont{Kaloper}}, \bibnamefont{and}
  \bibinfo{author}{\bibfnamefont{J.}~\bibnamefont{March-Russell}},
  \bibinfo{journal}{Physical Review D} \textbf{\bibinfo{volume}{81}},
  \bibinfo{pages}{123530} (\bibinfo{year}{2010}).

\bibitem[{\citenamefont{Green et~al.}(1987)\citenamefont{Green, Schwarz, and
  Witten}}]{Green:1987sp}
\bibinfo{author}{\bibfnamefont{M.~B.} \bibnamefont{Green}},
  \bibinfo{author}{\bibfnamefont{J.~H.} \bibnamefont{Schwarz}},
  \bibnamefont{and} \bibinfo{author}{\bibfnamefont{E.}~\bibnamefont{Witten}},
  \emph{\bibinfo{title}{{Superstring Theory}}}, Cambridge Monographs on
  Mathematical Physics (\bibinfo{year}{1987}).

\bibitem[{\citenamefont{Svrcek and Witten}(2006)}]{Svrcek:2006yi}
\bibinfo{author}{\bibfnamefont{P.}~\bibnamefont{Svrcek}} \bibnamefont{and}
  \bibinfo{author}{\bibfnamefont{E.}~\bibnamefont{Witten}},
  \bibinfo{journal}{JHEP} \textbf{\bibinfo{volume}{06}}, \bibinfo{pages}{051}
  (\bibinfo{year}{2006}), \eprint{hep-th/0605206}.

\bibitem[{\citenamefont{Arvanitaki and Dubovsky}(2011)}]{Arvanitaki:2010sy}
\bibinfo{author}{\bibfnamefont{A.}~\bibnamefont{Arvanitaki}} \bibnamefont{and}
  \bibinfo{author}{\bibfnamefont{S.}~\bibnamefont{Dubovsky}},
  \bibinfo{journal}{Phys. Rev.} \textbf{\bibinfo{volume}{D83}},
  \bibinfo{pages}{044026} (\bibinfo{year}{2011}), \eprint{1004.3558}.

\bibitem[{\citenamefont{Baumann et~al.}(2016)\citenamefont{Baumann, Green, and
  Wallisch}}]{Baumann:2016wac}
\bibinfo{author}{\bibfnamefont{D.}~\bibnamefont{Baumann}},
  \bibinfo{author}{\bibfnamefont{D.}~\bibnamefont{Green}}, \bibnamefont{and}
  \bibinfo{author}{\bibfnamefont{B.}~\bibnamefont{Wallisch}},
  \bibinfo{journal}{Phys. Rev. Lett.} \textbf{\bibinfo{volume}{117}},
  \bibinfo{pages}{171301} (\bibinfo{year}{2016}), \eprint{1604.08614}.

\bibitem[{\citenamefont{Rosa and Kephart}(2018)}]{Rosa:2017ury}
\bibinfo{author}{\bibfnamefont{J.~G.} \bibnamefont{Rosa}} \bibnamefont{and}
  \bibinfo{author}{\bibfnamefont{T.~W.} \bibnamefont{Kephart}},
  \bibinfo{journal}{Phys. Rev. Lett.} \textbf{\bibinfo{volume}{120}},
  \bibinfo{pages}{231102} (\bibinfo{year}{2018}), \eprint{1709.06581}.

\bibitem[{\citenamefont{Sagunski et~al.}(2018)\citenamefont{Sagunski, Zhang,
  Johnson, Lehner, Sakellariadou, Liebling, Palenzuela, and
  Neilsen}}]{Sagunski:2017nzb}
\bibinfo{author}{\bibfnamefont{L.}~\bibnamefont{Sagunski}},
  \bibinfo{author}{\bibfnamefont{J.}~\bibnamefont{Zhang}},
  \bibinfo{author}{\bibfnamefont{M.~C.} \bibnamefont{Johnson}},
  \bibinfo{author}{\bibfnamefont{L.}~\bibnamefont{Lehner}},
  \bibinfo{author}{\bibfnamefont{M.}~\bibnamefont{Sakellariadou}},
  \bibinfo{author}{\bibfnamefont{S.~L.} \bibnamefont{Liebling}},
  \bibinfo{author}{\bibfnamefont{C.}~\bibnamefont{Palenzuela}},
  \bibnamefont{and} \bibinfo{author}{\bibfnamefont{D.}~\bibnamefont{Neilsen}},
  \bibinfo{journal}{Phys. Rev.} \textbf{\bibinfo{volume}{D97}},
  \bibinfo{pages}{064016} (\bibinfo{year}{2018}), \eprint{1709.06634}.

\bibitem[{\citenamefont{Fujita et~al.}(2019)\citenamefont{Fujita, Tazaki, and
  Toma}}]{Fujita:2018zaj}
\bibinfo{author}{\bibfnamefont{T.}~\bibnamefont{Fujita}},
  \bibinfo{author}{\bibfnamefont{R.}~\bibnamefont{Tazaki}}, \bibnamefont{and}
  \bibinfo{author}{\bibfnamefont{K.}~\bibnamefont{Toma}},
  \bibinfo{journal}{Phys. Rev. Lett.} \textbf{\bibinfo{volume}{122}},
  \bibinfo{pages}{191101} (\bibinfo{year}{2019}), \eprint{1811.03525}.

\bibitem[{\citenamefont{Huang et~al.}(2019)\citenamefont{Huang, Johnson,
  Sagunski, Sakellariadou, and Zhang}}]{Huang:2018pbu}
\bibinfo{author}{\bibfnamefont{J.}~\bibnamefont{Huang}},
  \bibinfo{author}{\bibfnamefont{M.~C.} \bibnamefont{Johnson}},
  \bibinfo{author}{\bibfnamefont{L.}~\bibnamefont{Sagunski}},
  \bibinfo{author}{\bibfnamefont{M.}~\bibnamefont{Sakellariadou}},
  \bibnamefont{and} \bibinfo{author}{\bibfnamefont{J.}~\bibnamefont{Zhang}},
  \bibinfo{journal}{Phys. Rev.} \textbf{\bibinfo{volume}{D99}},
  \bibinfo{pages}{063013} (\bibinfo{year}{2019}), \eprint{1807.02133}.

\bibitem[{\citenamefont{Ikeda et~al.}(2019)\citenamefont{Ikeda, Brito, and
  Cardoso}}]{Ikeda:2019fvj}
\bibinfo{author}{\bibfnamefont{T.}~\bibnamefont{Ikeda}},
  \bibinfo{author}{\bibfnamefont{R.}~\bibnamefont{Brito}}, \bibnamefont{and}
  \bibinfo{author}{\bibfnamefont{V.}~\bibnamefont{Cardoso}},
  \bibinfo{journal}{Phys. Rev. Lett.} \textbf{\bibinfo{volume}{122}},
  \bibinfo{pages}{081101} (\bibinfo{year}{2019}), \eprint{1811.04950}.

\bibitem[{\citenamefont{Detweiler}(1980)}]{Detweiler:1980uk}
\bibinfo{author}{\bibfnamefont{S.~L.} \bibnamefont{Detweiler}},
  \bibinfo{journal}{Phys. Rev.} \textbf{\bibinfo{volume}{D22}},
  \bibinfo{pages}{2323} (\bibinfo{year}{1980}).

\bibitem[{\citenamefont{Zouros and Eardley}(1979)}]{zouros1979instabilities}
\bibinfo{author}{\bibfnamefont{T.~J.} \bibnamefont{Zouros}} \bibnamefont{and}
  \bibinfo{author}{\bibfnamefont{D.~M.} \bibnamefont{Eardley}},
  \bibinfo{journal}{Annals of physics} \textbf{\bibinfo{volume}{118}},
  \bibinfo{pages}{139} (\bibinfo{year}{1979}).

\bibitem[{\citenamefont{Brito et~al.}(2015)\citenamefont{Brito, Cardoso, and
  Pani}}]{Brito:2015oca}
\bibinfo{author}{\bibfnamefont{R.}~\bibnamefont{Brito}},
  \bibinfo{author}{\bibfnamefont{V.}~\bibnamefont{Cardoso}}, \bibnamefont{and}
  \bibinfo{author}{\bibfnamefont{P.}~\bibnamefont{Pani}},
  \bibinfo{journal}{Lect. Notes Phys.} \textbf{\bibinfo{volume}{906}},
  \bibinfo{pages}{pp.1} (\bibinfo{year}{2015}), \eprint{1501.06570}.

\bibitem[{\citenamefont{Dolan}(2007)}]{Dolan:2007mj}
\bibinfo{author}{\bibfnamefont{S.~R.} \bibnamefont{Dolan}},
  \bibinfo{journal}{Phys. Rev.} \textbf{\bibinfo{volume}{D76}},
  \bibinfo{pages}{084001} (\bibinfo{year}{2007}), \eprint{0705.2880}.

\bibitem[{\citenamefont{East and Pretorius}(2017)}]{east2017superradiant}
\bibinfo{author}{\bibfnamefont{W.~E.} \bibnamefont{East}} \bibnamefont{and}
  \bibinfo{author}{\bibfnamefont{F.}~\bibnamefont{Pretorius}},
  \bibinfo{journal}{Physical review letters} \textbf{\bibinfo{volume}{119}},
  \bibinfo{pages}{041101} (\bibinfo{year}{2017}).

\bibitem[{\citenamefont{East}(2017)}]{east2017superradiant2}
\bibinfo{author}{\bibfnamefont{W.~E.} \bibnamefont{East}},
  \bibinfo{journal}{Physical Review D} \textbf{\bibinfo{volume}{96}},
  \bibinfo{pages}{024004} (\bibinfo{year}{2017}).

\bibitem[{\citenamefont{Yoshino and Kodama}(2014)}]{Yoshino:2013ofa}
\bibinfo{author}{\bibfnamefont{H.}~\bibnamefont{Yoshino}} \bibnamefont{and}
  \bibinfo{author}{\bibfnamefont{H.}~\bibnamefont{Kodama}},
  \bibinfo{journal}{PTEP} \textbf{\bibinfo{volume}{2014}},
  \bibinfo{pages}{043E02} (\bibinfo{year}{2014}), \eprint{1312.2326}.

\bibitem[{\citenamefont{Arvanitaki et~al.}(2015)\citenamefont{Arvanitaki,
  Baryakhtar, and Huang}}]{arvanitaki2015discovering}
\bibinfo{author}{\bibfnamefont{A.}~\bibnamefont{Arvanitaki}},
  \bibinfo{author}{\bibfnamefont{M.}~\bibnamefont{Baryakhtar}},
  \bibnamefont{and} \bibinfo{author}{\bibfnamefont{X.}~\bibnamefont{Huang}},
  \bibinfo{journal}{Physical Review D} \textbf{\bibinfo{volume}{91}},
  \bibinfo{pages}{084011} (\bibinfo{year}{2015}).

\bibitem[{\citenamefont{Baryakhtar et~al.}(2017)\citenamefont{Baryakhtar,
  Lasenby, and Teo}}]{baryakhtar2017black}
\bibinfo{author}{\bibfnamefont{M.}~\bibnamefont{Baryakhtar}},
  \bibinfo{author}{\bibfnamefont{R.}~\bibnamefont{Lasenby}}, \bibnamefont{and}
  \bibinfo{author}{\bibfnamefont{M.}~\bibnamefont{Teo}},
  \bibinfo{journal}{Physical Review D} \textbf{\bibinfo{volume}{96}},
  \bibinfo{pages}{035019} (\bibinfo{year}{2017}).

\bibitem[{\citenamefont{Isi et~al.}(2019)\citenamefont{Isi, Sun, Brito, and
  Melatos}}]{isi2019directed}
\bibinfo{author}{\bibfnamefont{M.}~\bibnamefont{Isi}},
  \bibinfo{author}{\bibfnamefont{L.}~\bibnamefont{Sun}},
  \bibinfo{author}{\bibfnamefont{R.}~\bibnamefont{Brito}}, \bibnamefont{and}
  \bibinfo{author}{\bibfnamefont{A.}~\bibnamefont{Melatos}},
  \bibinfo{journal}{Physical Review D} \textbf{\bibinfo{volume}{99}},
  \bibinfo{pages}{084042} (\bibinfo{year}{2019}).

\bibitem[{\citenamefont{Yang et~al.}(2017)\citenamefont{Yang, Yagi, Blackman,
  Lehner, Paschalidis, Pretorius, and Yunes}}]{yang2017black}
\bibinfo{author}{\bibfnamefont{H.}~\bibnamefont{Yang}},
  \bibinfo{author}{\bibfnamefont{K.}~\bibnamefont{Yagi}},
  \bibinfo{author}{\bibfnamefont{J.}~\bibnamefont{Blackman}},
  \bibinfo{author}{\bibfnamefont{L.}~\bibnamefont{Lehner}},
  \bibinfo{author}{\bibfnamefont{V.}~\bibnamefont{Paschalidis}},
  \bibinfo{author}{\bibfnamefont{F.}~\bibnamefont{Pretorius}},
  \bibnamefont{and} \bibinfo{author}{\bibfnamefont{N.}~\bibnamefont{Yunes}},
  \bibinfo{journal}{Physical review letters} \textbf{\bibinfo{volume}{118}},
  \bibinfo{pages}{161101} (\bibinfo{year}{2017}).

\bibitem[{\citenamefont{Goncharov and Thrane}(2018)}]{goncharov2018all}
\bibinfo{author}{\bibfnamefont{B.}~\bibnamefont{Goncharov}} \bibnamefont{and}
  \bibinfo{author}{\bibfnamefont{E.}~\bibnamefont{Thrane}},
  \bibinfo{journal}{Physical Review D} \textbf{\bibinfo{volume}{98}},
  \bibinfo{pages}{064018} (\bibinfo{year}{2018}).

\bibitem[{\citenamefont{Pierce et~al.}(2018)\citenamefont{Pierce, Riles, and
  Zhao}}]{pierce2018searching}
\bibinfo{author}{\bibfnamefont{A.}~\bibnamefont{Pierce}},
  \bibinfo{author}{\bibfnamefont{K.}~\bibnamefont{Riles}}, \bibnamefont{and}
  \bibinfo{author}{\bibfnamefont{Y.}~\bibnamefont{Zhao}},
  \bibinfo{journal}{Physical review letters} \textbf{\bibinfo{volume}{121}},
  \bibinfo{pages}{061102} (\bibinfo{year}{2018}).

\bibitem[{\citenamefont{Brito et~al.}(2017{\natexlab{a}})\citenamefont{Brito,
  Ghosh, Barausse, Berti, Cardoso, Dvorkin, Klein, and Pani}}]{Brito:2017zvb}
\bibinfo{author}{\bibfnamefont{R.}~\bibnamefont{Brito}},
  \bibinfo{author}{\bibfnamefont{S.}~\bibnamefont{Ghosh}},
  \bibinfo{author}{\bibfnamefont{E.}~\bibnamefont{Barausse}},
  \bibinfo{author}{\bibfnamefont{E.}~\bibnamefont{Berti}},
  \bibinfo{author}{\bibfnamefont{V.}~\bibnamefont{Cardoso}},
  \bibinfo{author}{\bibfnamefont{I.}~\bibnamefont{Dvorkin}},
  \bibinfo{author}{\bibfnamefont{A.}~\bibnamefont{Klein}}, \bibnamefont{and}
  \bibinfo{author}{\bibfnamefont{P.}~\bibnamefont{Pani}},
  \bibinfo{journal}{Phys. Rev.} \textbf{\bibinfo{volume}{D96}},
  \bibinfo{pages}{064050} (\bibinfo{year}{2017}{\natexlab{a}}),
  \eprint{1706.06311}.

\bibitem[{\citenamefont{Brito et~al.}(2017{\natexlab{b}})\citenamefont{Brito,
  Ghosh, Barausse, Berti, Cardoso, Dvorkin, Klein, and Pani}}]{Brito:2017wnc}
\bibinfo{author}{\bibfnamefont{R.}~\bibnamefont{Brito}},
  \bibinfo{author}{\bibfnamefont{S.}~\bibnamefont{Ghosh}},
  \bibinfo{author}{\bibfnamefont{E.}~\bibnamefont{Barausse}},
  \bibinfo{author}{\bibfnamefont{E.}~\bibnamefont{Berti}},
  \bibinfo{author}{\bibfnamefont{V.}~\bibnamefont{Cardoso}},
  \bibinfo{author}{\bibfnamefont{I.}~\bibnamefont{Dvorkin}},
  \bibinfo{author}{\bibfnamefont{A.}~\bibnamefont{Klein}}, \bibnamefont{and}
  \bibinfo{author}{\bibfnamefont{P.}~\bibnamefont{Pani}},
  \bibinfo{journal}{Phys. Rev. Lett.} \textbf{\bibinfo{volume}{119}},
  \bibinfo{pages}{131101} (\bibinfo{year}{2017}{\natexlab{b}}),
  \eprint{1706.05097}.

\bibitem[{\citenamefont{Baumann
  et~al.}(2019{\natexlab{a}})\citenamefont{Baumann, Chia, and
  Porto}}]{Baumann:2018vus}
\bibinfo{author}{\bibfnamefont{D.}~\bibnamefont{Baumann}},
  \bibinfo{author}{\bibfnamefont{H.~S.} \bibnamefont{Chia}}, \bibnamefont{and}
  \bibinfo{author}{\bibfnamefont{R.~A.} \bibnamefont{Porto}},
  \bibinfo{journal}{Phys. Rev.} \textbf{\bibinfo{volume}{D99}},
  \bibinfo{pages}{044001} (\bibinfo{year}{2019}{\natexlab{a}}),
  \eprint{1804.03208}.

\bibitem[{\citenamefont{Berti et~al.}(2019)\citenamefont{Berti, Brito, Macedo,
  Raposo, and Rosa}}]{Berti:2019wnn}
\bibinfo{author}{\bibfnamefont{E.}~\bibnamefont{Berti}},
  \bibinfo{author}{\bibfnamefont{R.}~\bibnamefont{Brito}},
  \bibinfo{author}{\bibfnamefont{C.~F.~B.} \bibnamefont{Macedo}},
  \bibinfo{author}{\bibfnamefont{G.}~\bibnamefont{Raposo}}, \bibnamefont{and}
  \bibinfo{author}{\bibfnamefont{J.~L.} \bibnamefont{Rosa}},
  \bibinfo{journal}{Phys. Rev.} \textbf{\bibinfo{volume}{D99}},
  \bibinfo{pages}{104039} (\bibinfo{year}{2019}), \eprint{1904.03131}.

\bibitem[{\citenamefont{Ferreira et~al.}(2017)\citenamefont{Ferreira, Macedo,
  and Cardoso}}]{Ferreira:2017pth}
\bibinfo{author}{\bibfnamefont{M.~C.} \bibnamefont{Ferreira}},
  \bibinfo{author}{\bibfnamefont{C.~F.~B.} \bibnamefont{Macedo}},
  \bibnamefont{and} \bibinfo{author}{\bibfnamefont{V.}~\bibnamefont{Cardoso}},
  \bibinfo{journal}{Phys. Rev.} \textbf{\bibinfo{volume}{D96}},
  \bibinfo{pages}{083017} (\bibinfo{year}{2017}), \eprint{1710.00830}.

\bibitem[{\citenamefont{Hannuksela et~al.}(2019)\citenamefont{Hannuksela, Wong,
  Brito, Berti, and Li}}]{Hannuksela:2018izj}
\bibinfo{author}{\bibfnamefont{O.~A.} \bibnamefont{Hannuksela}},
  \bibinfo{author}{\bibfnamefont{K.~W.~K.} \bibnamefont{Wong}},
  \bibinfo{author}{\bibfnamefont{R.}~\bibnamefont{Brito}},
  \bibinfo{author}{\bibfnamefont{E.}~\bibnamefont{Berti}}, \bibnamefont{and}
  \bibinfo{author}{\bibfnamefont{T.~G.~F.} \bibnamefont{Li}},
  \bibinfo{journal}{Nat. Astron.} \textbf{\bibinfo{volume}{3}},
  \bibinfo{pages}{447} (\bibinfo{year}{2019}), \eprint{1804.09659}.

\bibitem[{\citenamefont{Zhang and Yang}(2019)}]{Zhang:2018kib}
\bibinfo{author}{\bibfnamefont{J.}~\bibnamefont{Zhang}} \bibnamefont{and}
  \bibinfo{author}{\bibfnamefont{H.}~\bibnamefont{Yang}},
  \bibinfo{journal}{Phys. Rev.} \textbf{\bibinfo{volume}{D99}},
  \bibinfo{pages}{064018} (\bibinfo{year}{2019}), \eprint{1808.02905}.

\bibitem[{\citenamefont{Ostriker}(1999)}]{Ostriker:1998fa}
\bibinfo{author}{\bibfnamefont{E.~C.} \bibnamefont{Ostriker}},
  \bibinfo{journal}{Astrophys. J.} \textbf{\bibinfo{volume}{513}},
  \bibinfo{pages}{252} (\bibinfo{year}{1999}), \eprint{astro-ph/9810324}.

\bibitem[{\citenamefont{Chandrasekhar}(1943)}]{Chandrasekhar:1943ys}
\bibinfo{author}{\bibfnamefont{S.}~\bibnamefont{Chandrasekhar}},
  \bibinfo{journal}{Astrophys. J.} \textbf{\bibinfo{volume}{97}},
  \bibinfo{pages}{255} (\bibinfo{year}{1943}).

\bibitem[{\citenamefont{Peters and Mathews}(1963)}]{Peters:1963ux}
\bibinfo{author}{\bibfnamefont{P.~C.} \bibnamefont{Peters}} \bibnamefont{and}
  \bibinfo{author}{\bibfnamefont{J.}~\bibnamefont{Mathews}},
  \bibinfo{journal}{Phys. Rev.} \textbf{\bibinfo{volume}{131}},
  \bibinfo{pages}{435} (\bibinfo{year}{1963}).

\bibitem[{\citenamefont{Benone et~al.}(2014)\citenamefont{Benone, de~Oliveira,
  Dolan, and Crispino}}]{Benone:2014qaa}
\bibinfo{author}{\bibfnamefont{C.~L.} \bibnamefont{Benone}},
  \bibinfo{author}{\bibfnamefont{E.~S.} \bibnamefont{de~Oliveira}},
  \bibinfo{author}{\bibfnamefont{S.~R.} \bibnamefont{Dolan}}, \bibnamefont{and}
  \bibinfo{author}{\bibfnamefont{L.~C.~B.} \bibnamefont{Crispino}},
  \bibinfo{journal}{Phys. Rev.} \textbf{\bibinfo{volume}{D89}},
  \bibinfo{pages}{104053} (\bibinfo{year}{2014}), \eprint{1404.0687}.

\bibitem[{\citenamefont{Benone et~al.}(2017)\citenamefont{Benone, de~Oliveira,
  Dolan, and Crispino}}]{Benone:2017xmg}
\bibinfo{author}{\bibfnamefont{C.~L.} \bibnamefont{Benone}},
  \bibinfo{author}{\bibfnamefont{E.~S.} \bibnamefont{de~Oliveira}},
  \bibinfo{author}{\bibfnamefont{S.~R.} \bibnamefont{Dolan}}, \bibnamefont{and}
  \bibinfo{author}{\bibfnamefont{L.~C.~B.} \bibnamefont{Crispino}},
  \bibinfo{journal}{Phys. Rev.} \textbf{\bibinfo{volume}{D95}},
  \bibinfo{pages}{044035} (\bibinfo{year}{2017}), \eprint{1702.06591}.

\bibitem[{\citenamefont{Unruh}(1976)}]{Unruh:1976fm}
\bibinfo{author}{\bibfnamefont{W.~G.} \bibnamefont{Unruh}},
  \bibinfo{journal}{Phys. Rev.} \textbf{\bibinfo{volume}{D14}},
  \bibinfo{pages}{3251} (\bibinfo{year}{1976}).

\bibitem[{\citenamefont{Press and Teukolsky}(1977)}]{press1977formation}
\bibinfo{author}{\bibfnamefont{W.}~\bibnamefont{Press}} \bibnamefont{and}
  \bibinfo{author}{\bibfnamefont{S.}~\bibnamefont{Teukolsky}},
  \bibinfo{journal}{The Astrophysical Journal} \textbf{\bibinfo{volume}{213}},
  \bibinfo{pages}{183} (\bibinfo{year}{1977}).

\bibitem[{\citenamefont{Yang et~al.}(2018)\citenamefont{Yang, East,
  Paschalidis, Pretorius, and Mendes}}]{yang2018evolution}
\bibinfo{author}{\bibfnamefont{H.}~\bibnamefont{Yang}},
  \bibinfo{author}{\bibfnamefont{W.~E.} \bibnamefont{East}},
  \bibinfo{author}{\bibfnamefont{V.}~\bibnamefont{Paschalidis}},
  \bibinfo{author}{\bibfnamefont{F.}~\bibnamefont{Pretorius}},
  \bibnamefont{and} \bibinfo{author}{\bibfnamefont{R.~F.}
  \bibnamefont{Mendes}}, \bibinfo{journal}{Physical Review D}
  \textbf{\bibinfo{volume}{98}}, \bibinfo{pages}{044007}
  (\bibinfo{year}{2018}).

\bibitem[{\citenamefont{Yang}(2019)}]{yang2019inspiraling}
\bibinfo{author}{\bibfnamefont{H.}~\bibnamefont{Yang}}, \bibinfo{journal}{arXiv
  preprint arXiv:1904.11089}  (\bibinfo{year}{2019}).

\bibitem[{\citenamefont{Mark et~al.}(2015)\citenamefont{Mark, Yang, Zimmerman,
  and Chen}}]{mark2015quasinormal}
\bibinfo{author}{\bibfnamefont{Z.}~\bibnamefont{Mark}},
  \bibinfo{author}{\bibfnamefont{H.}~\bibnamefont{Yang}},
  \bibinfo{author}{\bibfnamefont{A.}~\bibnamefont{Zimmerman}},
  \bibnamefont{and} \bibinfo{author}{\bibfnamefont{Y.}~\bibnamefont{Chen}},
  \bibinfo{journal}{Physical Review D} \textbf{\bibinfo{volume}{91}},
  \bibinfo{pages}{044025} (\bibinfo{year}{2015}).

\bibitem[{\citenamefont{Yang et~al.}(2015)\citenamefont{Yang, Zimmerman, and
  Lehner}}]{yang2015turbulent}
\bibinfo{author}{\bibfnamefont{H.}~\bibnamefont{Yang}},
  \bibinfo{author}{\bibfnamefont{A.}~\bibnamefont{Zimmerman}},
  \bibnamefont{and} \bibinfo{author}{\bibfnamefont{L.}~\bibnamefont{Lehner}},
  \bibinfo{journal}{Physical review letters} \textbf{\bibinfo{volume}{114}},
  \bibinfo{pages}{081101} (\bibinfo{year}{2015}).

\bibitem[{\citenamefont{Yang and Zhang}(2016)}]{yang2016plasma}
\bibinfo{author}{\bibfnamefont{H.}~\bibnamefont{Yang}} \bibnamefont{and}
  \bibinfo{author}{\bibfnamefont{F.}~\bibnamefont{Zhang}},
  \bibinfo{journal}{The Astrophysical Journal} \textbf{\bibinfo{volume}{817}},
  \bibinfo{pages}{183} (\bibinfo{year}{2016}).

\bibitem[{\citenamefont{Berry et~al.}(2019)\citenamefont{Berry, Hughes,
  Sopuerta, Chua, Heffernan, Holley-Bockelmann, Mihaylov, Miller, and
  Sesana}}]{berry2019unique}
\bibinfo{author}{\bibfnamefont{C.~P.} \bibnamefont{Berry}},
  \bibinfo{author}{\bibfnamefont{S.~A.} \bibnamefont{Hughes}},
  \bibinfo{author}{\bibfnamefont{C.~F.} \bibnamefont{Sopuerta}},
  \bibinfo{author}{\bibfnamefont{A.~J.} \bibnamefont{Chua}},
  \bibinfo{author}{\bibfnamefont{A.}~\bibnamefont{Heffernan}},
  \bibinfo{author}{\bibfnamefont{K.}~\bibnamefont{Holley-Bockelmann}},
  \bibinfo{author}{\bibfnamefont{D.~P.} \bibnamefont{Mihaylov}},
  \bibinfo{author}{\bibfnamefont{M.~C.} \bibnamefont{Miller}},
  \bibnamefont{and} \bibinfo{author}{\bibfnamefont{A.}~\bibnamefont{Sesana}},
  \bibinfo{journal}{arXiv preprint arXiv:1903.03686}  (\bibinfo{year}{2019}).

\bibitem[{\citenamefont{{Merritt}}(2013)}]{Merritt:2013cqg}
\bibinfo{author}{\bibfnamefont{D.}~\bibnamefont{{Merritt}}},
  \bibinfo{journal}{Classical and Quantum Gravity}
  \textbf{\bibinfo{volume}{30}}, \bibinfo{eid}{244005} (\bibinfo{year}{2013}),
  \eprint{1307.3268}.

\bibitem[{\citenamefont{Hopman and Alexander}(2005)}]{Hopman:2005vr}
\bibinfo{author}{\bibfnamefont{C.}~\bibnamefont{Hopman}} \bibnamefont{and}
  \bibinfo{author}{\bibfnamefont{T.}~\bibnamefont{Alexander}},
  \bibinfo{journal}{Astrophys. J.} \textbf{\bibinfo{volume}{629}},
  \bibinfo{pages}{362} (\bibinfo{year}{2005}), \eprint{astro-ph/0503672}.

\bibitem[{\citenamefont{Freitag and Benz}(2002)}]{Freitag:2002mj}
\bibinfo{author}{\bibfnamefont{M.}~\bibnamefont{Freitag}} \bibnamefont{and}
  \bibinfo{author}{\bibfnamefont{W.}~\bibnamefont{Benz}},
  \bibinfo{journal}{Astron. Astrophys.} \textbf{\bibinfo{volume}{394}},
  \bibinfo{pages}{345} (\bibinfo{year}{2002}), \eprint{astro-ph/0204292}.

\bibitem[{\citenamefont{Baumgardt
  et~al.}(2004{\natexlab{a}})\citenamefont{Baumgardt, Makino, and
  Ebisuzaki}}]{Baumgardt:2004zq}
\bibinfo{author}{\bibfnamefont{H.}~\bibnamefont{Baumgardt}},
  \bibinfo{author}{\bibfnamefont{J.}~\bibnamefont{Makino}}, \bibnamefont{and}
  \bibinfo{author}{\bibfnamefont{T.}~\bibnamefont{Ebisuzaki}},
  \bibinfo{journal}{Astrophys. J.} \textbf{\bibinfo{volume}{613}},
  \bibinfo{pages}{1133} (\bibinfo{year}{2004}{\natexlab{a}}),
  \eprint{astro-ph/0406227}.

\bibitem[{\citenamefont{Baumgardt
  et~al.}(2004{\natexlab{b}})\citenamefont{Baumgardt, Makino, and
  Ebisuzaki}}]{Baumgardt:2004zu}
\bibinfo{author}{\bibfnamefont{H.}~\bibnamefont{Baumgardt}},
  \bibinfo{author}{\bibfnamefont{J.}~\bibnamefont{Makino}}, \bibnamefont{and}
  \bibinfo{author}{\bibfnamefont{T.}~\bibnamefont{Ebisuzaki}},
  \bibinfo{journal}{Astrophys. J.} \textbf{\bibinfo{volume}{613}},
  \bibinfo{pages}{1143} (\bibinfo{year}{2004}{\natexlab{b}}),
  \eprint{astro-ph/0406231}.

\bibitem[{\citenamefont{Preto et~al.}(2004)\citenamefont{Preto, Merritt, and
  Spurzem}}]{Preto:2004kd}
\bibinfo{author}{\bibfnamefont{M.}~\bibnamefont{Preto}},
  \bibinfo{author}{\bibfnamefont{D.}~\bibnamefont{Merritt}}, \bibnamefont{and}
  \bibinfo{author}{\bibfnamefont{R.}~\bibnamefont{Spurzem}},
  \bibinfo{journal}{Astrophys. J.} \textbf{\bibinfo{volume}{613}},
  \bibinfo{pages}{L109} (\bibinfo{year}{2004}), \eprint{astro-ph/0406324}.

\bibitem[{\citenamefont{Alexander and Hopman}(2003)}]{Alexander:2003nc}
\bibinfo{author}{\bibfnamefont{T.}~\bibnamefont{Alexander}} \bibnamefont{and}
  \bibinfo{author}{\bibfnamefont{C.}~\bibnamefont{Hopman}},
  \bibinfo{journal}{Astrophys. J.} \textbf{\bibinfo{volume}{590}},
  \bibinfo{pages}{L29} (\bibinfo{year}{2003}), \eprint{astro-ph/0305062}.

\bibitem[{\citenamefont{Sesana}(2016)}]{sesana2016prospects}
\bibinfo{author}{\bibfnamefont{A.}~\bibnamefont{Sesana}},
  \bibinfo{journal}{Physical Review Letters} \textbf{\bibinfo{volume}{116}},
  \bibinfo{pages}{231102} (\bibinfo{year}{2016}).

\bibitem[{\citenamefont{Vitale}(2016)}]{vitale2016multiband}
\bibinfo{author}{\bibfnamefont{S.}~\bibnamefont{Vitale}},
  \bibinfo{journal}{Physical review letters} \textbf{\bibinfo{volume}{117}},
  \bibinfo{pages}{051102} (\bibinfo{year}{2016}).

\bibitem[{\citenamefont{Wong et~al.}(2018)\citenamefont{Wong, Kovetz, Cutler,
  and Berti}}]{wong2018expanding}
\bibinfo{author}{\bibfnamefont{K.~W.} \bibnamefont{Wong}},
  \bibinfo{author}{\bibfnamefont{E.~D.} \bibnamefont{Kovetz}},
  \bibinfo{author}{\bibfnamefont{C.}~\bibnamefont{Cutler}}, \bibnamefont{and}
  \bibinfo{author}{\bibfnamefont{E.}~\bibnamefont{Berti}},
  \bibinfo{journal}{Physical review letters} \textbf{\bibinfo{volume}{121}},
  \bibinfo{pages}{251102} (\bibinfo{year}{2018}).

\bibitem[{\citenamefont{Cutler et~al.}(2019)\citenamefont{Cutler, Berti, Jani,
  Kovetz, Randall, Vitale, Wong, Holley-Bockelmann, Larson, Littenberg
  et~al.}}]{cutler2019we}
\bibinfo{author}{\bibfnamefont{C.}~\bibnamefont{Cutler}},
  \bibinfo{author}{\bibfnamefont{E.}~\bibnamefont{Berti}},
  \bibinfo{author}{\bibfnamefont{K.}~\bibnamefont{Jani}},
  \bibinfo{author}{\bibfnamefont{E.~D.} \bibnamefont{Kovetz}},
  \bibinfo{author}{\bibfnamefont{L.}~\bibnamefont{Randall}},
  \bibinfo{author}{\bibfnamefont{S.}~\bibnamefont{Vitale}},
  \bibinfo{author}{\bibfnamefont{K.~W.} \bibnamefont{Wong}},
  \bibinfo{author}{\bibfnamefont{K.}~\bibnamefont{Holley-Bockelmann}},
  \bibinfo{author}{\bibfnamefont{S.~L.} \bibnamefont{Larson}},
  \bibinfo{author}{\bibfnamefont{T.}~\bibnamefont{Littenberg}},
  \bibnamefont{et~al.}, \bibinfo{journal}{arXiv preprint arXiv:1903.04069}
  (\bibinfo{year}{2019}).

\bibitem[{\citenamefont{Cutler and Flanagan}(1994)}]{Cutler:1994ys}
\bibinfo{author}{\bibfnamefont{C.}~\bibnamefont{Cutler}} \bibnamefont{and}
  \bibinfo{author}{\bibfnamefont{E.~E.} \bibnamefont{Flanagan}},
  \bibinfo{journal}{Phys. Rev.} \textbf{\bibinfo{volume}{D49}},
  \bibinfo{pages}{2658} (\bibinfo{year}{1994}), \eprint{gr-qc/9402014}.

\bibitem[{\citenamefont{Flanagan and Hinderer}(2012)}]{flanagan2012transient}
\bibinfo{author}{\bibfnamefont{E.~E.} \bibnamefont{Flanagan}} \bibnamefont{and}
  \bibinfo{author}{\bibfnamefont{T.}~\bibnamefont{Hinderer}},
  \bibinfo{journal}{Physical review letters} \textbf{\bibinfo{volume}{109}},
  \bibinfo{pages}{071102} (\bibinfo{year}{2012}).

\bibitem[{\citenamefont{Yang and Casals}(2017)}]{yang2017general}
\bibinfo{author}{\bibfnamefont{H.}~\bibnamefont{Yang}} \bibnamefont{and}
  \bibinfo{author}{\bibfnamefont{M.}~\bibnamefont{Casals}},
  \bibinfo{journal}{Physical Review D} \textbf{\bibinfo{volume}{96}},
  \bibinfo{pages}{083015} (\bibinfo{year}{2017}).

\bibitem[{\citenamefont{Bonga et~al.}(2019)\citenamefont{Bonga, Yang, and
  Hughes}}]{bonga2019tidal}
\bibinfo{author}{\bibfnamefont{B.}~\bibnamefont{Bonga}},
  \bibinfo{author}{\bibfnamefont{H.}~\bibnamefont{Yang}}, \bibnamefont{and}
  \bibinfo{author}{\bibfnamefont{S.~A.} \bibnamefont{Hughes}},
  \bibinfo{journal}{arXiv preprint arXiv:1905.00030}  (\bibinfo{year}{2019}).

\bibitem[{\citenamefont{O'Leary et~al.}(2009)\citenamefont{O'Leary, Kocsis, and
  Loeb}}]{OLeary:2008myb}
\bibinfo{author}{\bibfnamefont{R.~M.} \bibnamefont{O'Leary}},
  \bibinfo{author}{\bibfnamefont{B.}~\bibnamefont{Kocsis}}, \bibnamefont{and}
  \bibinfo{author}{\bibfnamefont{A.}~\bibnamefont{Loeb}},
  \bibinfo{journal}{Mon. Not. Roy. Astron. Soc.}
  \textbf{\bibinfo{volume}{395}}, \bibinfo{pages}{2127} (\bibinfo{year}{2009}),
  \eprint{0807.2638}.

\bibitem[{\citenamefont{Barack et~al.}(2019)}]{Barack:2018yly}
\bibinfo{author}{\bibfnamefont{L.}~\bibnamefont{Barack}} \bibnamefont{et~al.}
  (\bibinfo{collaboration}{LIGO}), \bibinfo{journal}{Class. Quant. Grav.}
  \textbf{\bibinfo{volume}{36}}, \bibinfo{pages}{143001}
  (\bibinfo{year}{2019}), \eprint{1806.05195}.

\bibitem[{\citenamefont{Abbott et~al.}(2017)\citenamefont{Abbott, Abbott,
  Abbott, Abernathy, Ackley, Adams, Addesso, Adhikari, Adya, Affeldt
  et~al.}}]{abbott2017exploring}
\bibinfo{author}{\bibfnamefont{B.~P.} \bibnamefont{Abbott}},
  \bibinfo{author}{\bibfnamefont{R.}~\bibnamefont{Abbott}},
  \bibinfo{author}{\bibfnamefont{T.}~\bibnamefont{Abbott}},
  \bibinfo{author}{\bibfnamefont{M.}~\bibnamefont{Abernathy}},
  \bibinfo{author}{\bibfnamefont{K.}~\bibnamefont{Ackley}},
  \bibinfo{author}{\bibfnamefont{C.}~\bibnamefont{Adams}},
  \bibinfo{author}{\bibfnamefont{P.}~\bibnamefont{Addesso}},
  \bibinfo{author}{\bibfnamefont{R.}~\bibnamefont{Adhikari}},
  \bibinfo{author}{\bibfnamefont{V.}~\bibnamefont{Adya}},
  \bibinfo{author}{\bibfnamefont{C.}~\bibnamefont{Affeldt}},
  \bibnamefont{et~al.}, \bibinfo{journal}{Classical and Quantum Gravity}
  \textbf{\bibinfo{volume}{34}}, \bibinfo{pages}{044001}
  (\bibinfo{year}{2017}).

\bibitem[{\citenamefont{Belczynski et~al.}(2016)\citenamefont{Belczynski,
  Heger, Gladysz, Ruiter, Woosley, Wiktorowicz, Chen, Bulik, O?Shaughnessy,
  Holz et~al.}}]{belczynski2016effect}
\bibinfo{author}{\bibfnamefont{K.}~\bibnamefont{Belczynski}},
  \bibinfo{author}{\bibfnamefont{A.}~\bibnamefont{Heger}},
  \bibinfo{author}{\bibfnamefont{W.}~\bibnamefont{Gladysz}},
  \bibinfo{author}{\bibfnamefont{A.~J.} \bibnamefont{Ruiter}},
  \bibinfo{author}{\bibfnamefont{S.}~\bibnamefont{Woosley}},
  \bibinfo{author}{\bibfnamefont{G.}~\bibnamefont{Wiktorowicz}},
  \bibinfo{author}{\bibfnamefont{H.-Y.} \bibnamefont{Chen}},
  \bibinfo{author}{\bibfnamefont{T.}~\bibnamefont{Bulik}},
  \bibinfo{author}{\bibfnamefont{R.}~\bibnamefont{O?Shaughnessy}},
  \bibinfo{author}{\bibfnamefont{D.~E.} \bibnamefont{Holz}},
  \bibnamefont{et~al.}, \bibinfo{journal}{Astronomy \& Astrophysics}
  \textbf{\bibinfo{volume}{594}}, \bibinfo{pages}{A97} (\bibinfo{year}{2016}).

\bibitem[{\citenamefont{Spera and Mapelli}(2017)}]{spera2017very}
\bibinfo{author}{\bibfnamefont{M.}~\bibnamefont{Spera}} \bibnamefont{and}
  \bibinfo{author}{\bibfnamefont{M.}~\bibnamefont{Mapelli}},
  \bibinfo{journal}{Monthly Notices of the Royal Astronomical Society}
  \textbf{\bibinfo{volume}{470}}, \bibinfo{pages}{4739} (\bibinfo{year}{2017}).

\bibitem[{\citenamefont{Marchant et~al.}(2018)\citenamefont{Marchant, Renzo,
  Farmer, Pappas, Taam, de~Mink, and Kalogera}}]{marchant2018pulsational}
\bibinfo{author}{\bibfnamefont{P.}~\bibnamefont{Marchant}},
  \bibinfo{author}{\bibfnamefont{M.}~\bibnamefont{Renzo}},
  \bibinfo{author}{\bibfnamefont{R.}~\bibnamefont{Farmer}},
  \bibinfo{author}{\bibfnamefont{K.~M.} \bibnamefont{Pappas}},
  \bibinfo{author}{\bibfnamefont{R.~E.} \bibnamefont{Taam}},
  \bibinfo{author}{\bibfnamefont{S.}~\bibnamefont{de~Mink}}, \bibnamefont{and}
  \bibinfo{author}{\bibfnamefont{V.}~\bibnamefont{Kalogera}},
  \bibinfo{journal}{arXiv preprint arXiv:1810.13412}  (\bibinfo{year}{2018}).

\bibitem[{\citenamefont{Stevenson et~al.}(2019)\citenamefont{Stevenson,
  Sampson, Powell, Vigna-G{\'o}mez, Neijssel, Sz{\'e}csi, and
  Mandel}}]{stevenson2019impact}
\bibinfo{author}{\bibfnamefont{S.}~\bibnamefont{Stevenson}},
  \bibinfo{author}{\bibfnamefont{M.}~\bibnamefont{Sampson}},
  \bibinfo{author}{\bibfnamefont{J.}~\bibnamefont{Powell}},
  \bibinfo{author}{\bibfnamefont{A.}~\bibnamefont{Vigna-G{\'o}mez}},
  \bibinfo{author}{\bibfnamefont{C.~J.} \bibnamefont{Neijssel}},
  \bibinfo{author}{\bibfnamefont{D.}~\bibnamefont{Sz{\'e}csi}},
  \bibnamefont{and} \bibinfo{author}{\bibfnamefont{I.}~\bibnamefont{Mandel}},
  \bibinfo{journal}{arXiv preprint arXiv:1904.02821}  (\bibinfo{year}{2019}).

\bibitem[{\citenamefont{Woosley}(2017)}]{woosley2017pulsational}
\bibinfo{author}{\bibfnamefont{S.}~\bibnamefont{Woosley}},
  \bibinfo{journal}{The Astrophysical Journal} \textbf{\bibinfo{volume}{836}},
  \bibinfo{pages}{244} (\bibinfo{year}{2017}).

\bibitem[{\citenamefont{Gerosa and Berti}(2019)}]{gerosa2019escape}
\bibinfo{author}{\bibfnamefont{D.}~\bibnamefont{Gerosa}} \bibnamefont{and}
  \bibinfo{author}{\bibfnamefont{E.}~\bibnamefont{Berti}},
  \bibinfo{journal}{arXiv preprint arXiv:1906.05295}  (\bibinfo{year}{2019}).

\bibitem[{\citenamefont{Hopman and Alexander}(2006)}]{Hopman:2006qr}
\bibinfo{author}{\bibfnamefont{C.}~\bibnamefont{Hopman}} \bibnamefont{and}
  \bibinfo{author}{\bibfnamefont{T.}~\bibnamefont{Alexander}},
  \bibinfo{journal}{Astrophys. J.} \textbf{\bibinfo{volume}{645}},
  \bibinfo{pages}{1152} (\bibinfo{year}{2006}), \eprint{astro-ph/0601161}.

\bibitem[{\citenamefont{Bahcall and Wolf}(1976)}]{Bahcall:1976aa}
\bibinfo{author}{\bibfnamefont{J.~N.} \bibnamefont{Bahcall}} \bibnamefont{and}
  \bibinfo{author}{\bibfnamefont{R.~A.} \bibnamefont{Wolf}},
  \bibinfo{journal}{Astrophys. J.} \textbf{\bibinfo{volume}{209}},
  \bibinfo{pages}{214} (\bibinfo{year}{1976}).

\bibitem[{\citenamefont{Fan et~al.}(2003)\citenamefont{Fan, Strauss, Schneider,
  Becker, White, Haiman, Gregg, Pentericci, Grebel, Narayanan
  et~al.}}]{fan2003survey}
\bibinfo{author}{\bibfnamefont{X.}~\bibnamefont{Fan}},
  \bibinfo{author}{\bibfnamefont{M.~A.} \bibnamefont{Strauss}},
  \bibinfo{author}{\bibfnamefont{D.~P.} \bibnamefont{Schneider}},
  \bibinfo{author}{\bibfnamefont{R.~H.} \bibnamefont{Becker}},
  \bibinfo{author}{\bibfnamefont{R.~L.} \bibnamefont{White}},
  \bibinfo{author}{\bibfnamefont{Z.}~\bibnamefont{Haiman}},
  \bibinfo{author}{\bibfnamefont{M.}~\bibnamefont{Gregg}},
  \bibinfo{author}{\bibfnamefont{L.}~\bibnamefont{Pentericci}},
  \bibinfo{author}{\bibfnamefont{E.~K.} \bibnamefont{Grebel}},
  \bibinfo{author}{\bibfnamefont{V.~K.} \bibnamefont{Narayanan}},
  \bibnamefont{et~al.}, \bibinfo{journal}{The Astronomical Journal}
  \textbf{\bibinfo{volume}{125}}, \bibinfo{pages}{1649} (\bibinfo{year}{2003}).

\bibitem[{\citenamefont{Natarajan et~al.}(2019)\citenamefont{Natarajan,
  Ricarte, Baldassare, Bellovary, Bender, Berti, Cappelluti, Ferrara, Greene,
  Haiman et~al.}}]{natarajan2019disentangling}
\bibinfo{author}{\bibfnamefont{P.}~\bibnamefont{Natarajan}},
  \bibinfo{author}{\bibfnamefont{A.}~\bibnamefont{Ricarte}},
  \bibinfo{author}{\bibfnamefont{V.}~\bibnamefont{Baldassare}},
  \bibinfo{author}{\bibfnamefont{J.}~\bibnamefont{Bellovary}},
  \bibinfo{author}{\bibfnamefont{P.}~\bibnamefont{Bender}},
  \bibinfo{author}{\bibfnamefont{E.}~\bibnamefont{Berti}},
  \bibinfo{author}{\bibfnamefont{N.}~\bibnamefont{Cappelluti}},
  \bibinfo{author}{\bibfnamefont{A.}~\bibnamefont{Ferrara}},
  \bibinfo{author}{\bibfnamefont{J.}~\bibnamefont{Greene}},
  \bibinfo{author}{\bibfnamefont{Z.}~\bibnamefont{Haiman}},
  \bibnamefont{et~al.}, \bibinfo{journal}{arXiv preprint arXiv:1904.09326}
  (\bibinfo{year}{2019}).

\bibitem[{\citenamefont{Baumann
  et~al.}(2019{\natexlab{b}})\citenamefont{Baumann, Chia, Porto, and
  Stout}}]{Baumann:2019ztm}
\bibinfo{author}{\bibfnamefont{D.}~\bibnamefont{Baumann}},
  \bibinfo{author}{\bibfnamefont{H.~S.} \bibnamefont{Chia}},
  \bibinfo{author}{\bibfnamefont{R.~A.} \bibnamefont{Porto}}, \bibnamefont{and}
  \bibinfo{author}{\bibfnamefont{J.}~\bibnamefont{Stout}}
  (\bibinfo{year}{2019}{\natexlab{b}}), \eprint{1912.04932}.

\end{thebibliography}
\end{document}